\documentclass[12pt]{iopart}
\usepackage{iopams}
\usepackage{epsf}

\newcommand{\rs}{\rm\scriptscriptstyle}
\newcommand{\bq}{\boldsymbol {q}}
\newcommand{\bk}{\boldsymbol {k}}

\begin{document}

\title[PEM in the gaussian fluctuation approximation]
 {Investigation of thermodynamic properties of pseudospin-electron
  model in the gaussian fluctuation approximation}

\author{K V Tabunshchyk}

\address{Institute for Condensed Matter Physics
         of the National Academy of Sciences of Ukraine,
         1 Svientsitskii Str., UA--79011 Lviv, Ukraine}

\ead{tkir@icmp.lviv.ua}

\begin{abstract}
 A method of consideration of gaussian fluctuations of the effective mean field
within the framework of the GRPA scheme is applied to investigation of
thermodynamical properties of a pseudospin-electron model (PEM).
 The grand canonical potential, pseudospin mean value, as well as
the mean squares of fluctuations are calculated.
 Obtained results are compared with corresponding ones obtained by
other approximations.
 An influence of the gaussian fluctuations of mean
field on the thermodynamic properties of PEM is discussed.

\end{abstract}

\submitto{\JPCM}

\maketitle

\section{Introduction}
 The pseudospin-electron model (PEM) was originally proposed to include a local
interaction of the conducting electrons in metals (or semimetals) with a
certain two level subsystem when it is reasonable to use
pseudospin formalism; the pseudospin variable
$S_{i}^{z}=\pm 1/2$ defines these two states.
  Starting from \cite{Muller,Hirsch} this scheme is applied
to describe the strongly correlated electrons of CuO$_2$ sheets
coupled with the vibrational states of apex oxygen ions O$_{\rm IV}$
(moving in a double-well potential) in YBaCuO type systems.
 Recently a similar model was proposed for investigation of the
proton-electron interaction in molecular and crystalline systems
with hydrogen bonds \cite{Matsushita1,Matsushita2}.
 The model Hamiltonian has the following form:
\begin{eqnarray}
 \label{H_initial}
 H=\sum_iH_i+\sum_{ij\sigma}t_{ij}c^+_{i\sigma}c_{j\sigma},
\end{eqnarray}
and includes, in addition to the terms describing electron transfer ($\sim t_{ij}$),
the electron correlation ($U$ - term) in the spirit of the Hubbard model,
interaction with the anharmonic mode ($g$ - term),
the energy of the tunnelling splitting ($\Omega$ - term),
and the energy of the anharmonic potential asymmetry ($h$ - term)
in the single-site part
\begin{eqnarray}
\label{H_initial0}
\fl H_i=Un_{i\uparrow}n_{i\downarrow}-\mu(n_{i\uparrow}+n_{i\downarrow})+
    g(n_{i\uparrow}+n_{i\downarrow})S^z_i-\Omega S^x_i-hS^z_i.
\end{eqnarray}

 Based on PEM a possible connection between the
superconductivity and lattice instability of the ferroelectric
type in HTSC was discussed \cite{Hirsch,Frick}.
 Description of the electron spectrum and electron statistics of the PEM
was given in \cite{Stasyuk} within the framework of the temperature two-time
Green's function method in the Hubbard-I approximation.

 A series of works has been carried out in which the pseudospin
$\left\langle S^zS^z\right\rangle$, mixed $\left\langle
S^zn\right\rangle$, and charge  $\left\langle nn\right\rangle $
correlation functions were calculated.
 A possibility of divergences of these functions at some
values of temperature exists as it was shown with making use of
the generalized random phase approximation (GRPA) \cite{Izyumov}
in the limit of infinite single-site electron correlations
($U\to\infty$) \cite{Stasyuk2,Stasyuk3}.
 This effect was interpreted as a manifestation of dielectric instability or
ferroelectric type anomaly.
 The tendency to form the spatially modulated charge and pseudospin ordering
at certain model parameter values was established.

 The case of absence of electron transfer ($t_{ij}=0$), but with
taking into account of the direct interaction between pseudospins
was considered within the mean field approximation
\cite{Havrylyuk,Dublenych,Tabunshchyk4}.
 It was shown there that behaviour of the system strongly depends
on a thermodynamical regime: at $\mu={\rm const}$
the first order phase transitions with the jumps of
pseudospin mean value $\langle S^z\rangle$ and electron
concentration values $n$ as well as the second ones
(with smooth changes) were detected, while at $n={\rm const}$
an instability with respect to phase separation in the electron and
pseudospin subsystems can take place.

  In the papers \cite{Tabunshchyk1,Shvaika,Tabunshchyk2,Tabunshchyk3}
PEM was considered at the absence of direct electron-electron interaction
and tunnelling splitting ($U=0$, $\Omega=0$).
 In such a version of PEM (simplified PEM), an effective many-body pseudospin
interaction via conducting electrons appears, and hence it is
interesting to investigate an influence of the retarded nondirect
interaction between pseudospins on thermodynamics of the model.
 Hamiltonian of this simplified PEM is invariant with
respect to the transformation $\mu\rightarrow -\mu$,
$h\rightarrow 2g-h$, $n\rightarrow 2-n$, $S^z\rightarrow-S^z$
(the so-called electron-hole symmetry) what makes possible to
describe the hole-pseudospin system as well.

 On the other hand, the presented Hamiltonian in the case of
absence of electron correlation and tunnelling
splitting allows one to describe the binary alloy type model.
 It is convenient to introduce projective operators on pseudospin states
$P^{\pm}_i=1/2\pm S^z_i$; then Hamiltonian of the binary alloy can be obtained
by substitution $P^+_i=p_i$, $P^-_i=1-p_i$
where $p_i$ is the concentration of one component of binary alloy,
and $1-p_i$ is the concentration of the second one.
 Difference between these models is in the way how an
averaging procedure over projection operators $P^{\pm}$ is performed
(thermal statistical averaging in the case of PEM and configurational
averaging for binary alloy) and how the self-consistency is achieved
(fixed value of longitudinal field $h$ for PEM and fixed value of
the component concentration $p$ for binary alloy).

  Furthermore, if in Hamiltonian (\ref{H_initial}), (\ref{H_initial0})
(considering the simplified case $U=0$, $\Omega=0$),
we remove spin indices and rewrite the Hamiltonian in terms of the
operators of the mobile $d_i$ ($c_{i\sigma}=d_i$) and localized
$f_i$ ($P^+_i=f^+_if_i$, $P^-_i=1-f^+_if_i$) electrons, we obtain
the Hamiltonian of the Falicov-Kimball model
where $h$ plays a role of the chemical potential for the localized
$f$-electrons.
 The ground state of the Falicov-Kimball model, when the electron
concentration for subsystems is fixed, is not uniform and shows
the commensurate or incommensurate ordering or phase separation
depending on the concentration and coupling constant values
\cite{Freericks}.
 On the other hand, in the case of the fixed
value of field $h$, the bistability effects are observed.

 Investigations of the thermodynamic properties of the simplified PEM
within the framework of the self-consistent GRPA \cite{Tabunshchyk1}
and dynamical mean field approximation (DMFA) \cite{Shvaika} have shown that:
\begin{itemize}
 \item in the $\mu = {\rm const}$ regime (when the electron
states of other structure elements, which are not included
explicitly into the PEM, are supposed to play a role of a thermostat
ensuring a fixed value of the chemical potential $\mu$) the interaction
between the electron and pseudospin subsystems leads to the
possibility of either first or second order phase transitions
between different uniform phases (bistability effect)
\cite{Tabunshchyk1,Shvaika} as well as between
the uniform and the chess-board ones \cite{Tabunshchyk2,Tabunshchyk3};
 \item in the regime $n={\rm const}$ (this situation is more customary at
the consideration of electron systems and means that the
chemical potential depends now on the electron concentration
being the function of $T$, $h$ etc.) an instability with
respect to the phase separation in the electron and pseudospin
subsystems can take place \cite{Tabunshchyk1,Shvaika,Tabunshchyk3}.
\end{itemize}

 The above mentioned methods of investigation of the PEM
(within the framework of the self-consistent GRPA scheme as well as
within the framework of DMFA in the case of the finite space dimension)
take into account only mean field type contributions.
 It is reasonable only when fluctuation effects are small.
 In the vicinity of critical points effects of the mean field
fluctuations become significant.

 The aim of this article is to calculate the thermodynamic and correlation
functions of the simplified PEM ($U=0$, $\Omega=0$) within the framework of
self-consistent approach, allowing us take into account fluctuation
effects of the effective self-consistent mean field.
 Such a consistent gaussian fluctuation approach was proposed previously
in work \cite{Tabunshchyk} where expressions for the grand canonical potential and
pseudospin correlator as well as a self-consistent set of equations
for the pseudospin mean value and root--mean--squares (r.m.s.)
fluctuation parameter were obtained.

  Furthermore we would like to investigate an influence of the gaussian
fluctuations on the thermodynamics of phase transitions in the PEM and discuss
an applicability of the schemes previously used for the investigation of the PEM.
 For this purpose the results of investigation of PEM within the
framework of the self-consistent GRPA scheme \cite{Tabunshchyk1} as well as within
the DMFA \cite{Shvaika} are presented.

 The paper is organized as follows.
 First, a short review of the self-consistent GRPA approach is presented and
the results of the numerical calculations within the framework of this
scheme are compared with corresponding ones obtained within
the DMFA in \cite{Shvaika}.
 Second, we review the self-consistent gaussian fluctuation approach
for the PEM and present its simplified version (analogous to the approximation
proposed by Onyszkiewicz \cite{Onyshkevych1,Onyshkevych2} for spin models).
 Then we apply gaussian fluctuation method to investigation of the
thermodynamic properties of the PEM, and finally results of the
numerical calculations of the pseudospin mean value and r.m.s.
fluctuation parameter as well as the grand canonical potential are shown.
 Phase diagrams are presented and compared to the corresponding
ones of the GRPA.
 Finally, discussion and conclusion are given.

\section{Self-consistent GRPA method}

 We would like to remind that traditional GRPA method (proposed by
Izyumov  et al. \cite{Izyumov} for the investigation of the magnetic
susceptibility of the ordinary Hubbard model as well as $t-J$ model)
takes into account the same topological class of diagrams
(so-called loop diagrams) as in the random phase approximation (RPA).
 Main difference between the GRPA and RPA is a way of choosing of
the basic states: splitted (Hubbard-I) band states in the GRPA method
and pure band states in RPA.
 The question how to calculate thermodynamic quantities within the
GRPA has been open until recent works \cite{Tabunshchyk1,Danyliv}.
 In this section we briefly show how to construct the mean field type
approximation within the GRPA scheme for PEM.

 Calculations are performed in the strong coupling case ($g\gg t$) using
single-site states as the basic ones.
 The formalism of electron annihilation (creation) operators
$
 a_{i\sigma}=c_{i\sigma}P^+_i,
$
$
 \tilde{a}_{i\sigma}=c_{i\sigma}P^-_i
$
($P^{\pm}_i=1/2\pm S^z_i$) acting at a site with the certain
pseudospin orientation is introduced \cite{Tabunshchyk1}.
 Within the framework of the proposed representation,
the initial Hamiltonian (\ref{H_initial}) (in the case of $U=0$, $\Omega=0$)
has the following form:
\begin{eqnarray}
\label{Hamiltonian2}
\fl H=H_0{+}H_{\rm int}=\sum\limits_{i\sigma} \{
 \varepsilon_1        n_{i\sigma}{+}
 \varepsilon_2\tilde{n}_{i\sigma}{-}
 \frac h2 S^z_i\}+\sum\limits_{ij\sigma}t_{ij}
 (a^+_{i\sigma}a_{j\sigma}{+}
  a^+_{i\sigma}\tilde{a}_{j\sigma}{+}
 \tilde{a}^+_{i\sigma}a_{j\sigma}{+}
 \tilde{a}^+_{i\sigma}
 \tilde{a}_{j\sigma}),
\end{eqnarray}
where $\varepsilon_{1,2}=-\mu\pm g/2$
are the energies of the single-site states.

 Expansion of the calculated quantities in terms of the electron transfer
$H_{\rm int}$ leads to the infinite series of terms containing the averages of
the $T$-products of the $a_{i\sigma}$, $\tilde{a}_{i\sigma}$,
$a^+_{i\sigma}$, $\tilde{a}^+_{i\sigma}$
operators.
 The evaluation of such averages is made using the appropriate
Wick's theorem \cite{Tabunshchyk1} formulated in the spirit of
Wick's theorem for Hubbard operators \cite{Slobodyan}.
 This theorem gives an algorithm reducing the average of product
of $n$ creation (annihilation) operators to the sum of averages
of products of the $n-2$ operators.
 So it is possible to express the result in terms
of the products of nonperturbed Green's functions and averages
of the projection operators $P^{\pm}_i$ expanded in semi-invariants.

 Nonperturbated electron Green's function is equal to
\begin{eqnarray}
 \label{nonpert_green}
 g(\omega_n)=\langle g_i(\omega_n)\rangle;\quad
 g_i(\omega_n)=\frac{P^+_i}{\rmi\omega_n-\varepsilon_1}+
 \frac{P^-_i}{\rmi\omega_n-\varepsilon_2}.
\end{eqnarray}

 In the uniform case $\langle S^z_i \rangle=\langle S^z \rangle$, a
single-electron Green's function (calculated in Hubbard-I type approximation)
is
$
 \raisebox{-0.13cm}{\epsfysize .4cm\epsfbox{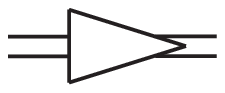}}
$
$
 =G_{\bk}(\omega_n)
$
$
=\big(g^{-1}(\omega_n)-t_{\bk}\big)^{-1}
$;
its poles determine the electron spectrum
\begin{eqnarray}
 \label{polus}
 \varepsilon_{\rs {I},\rs {II}}(t_{\bk})=
 \frac 12(-2\mu+t_{\bk})\pm \frac 12
 \sqrt{g^2+4t_{\bk}\langle S^z \rangle g +t^2_{\bk}}\, .
\end{eqnarray}
 Investigation of the electron spectrum was performed in
\cite{Tabunshchyk1}.
 In this approximation, the branches $\varepsilon_{\rs {I}}(t_{\bk})$ and
$\varepsilon_{\rs {II}}(t_{\bk})$ form two electron subbands
always separated by a gap.
 Reconstruction of the electron spectrum takes place
with change of the pseudospin mean value $\langle S^z \rangle$.

 In the adopted approximation the diagrammatic series for the pseudospin mean
value can be presented in the form \cite{Tabunshchyk1}
\begin{eqnarray}
 \label{Sz_diagr}
 \langle S^z\rangle= \raisebox{-.24cm}{\epsfysize 1.3cm\epsfbox{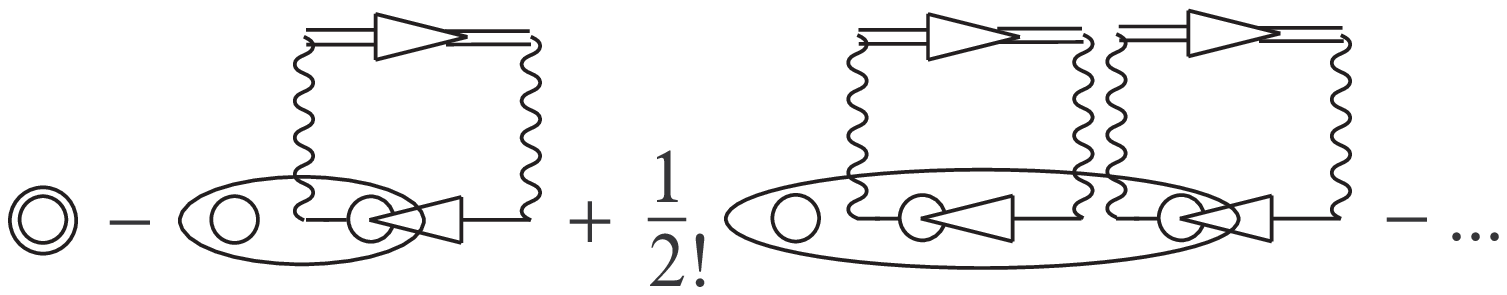}}\quad,
\end{eqnarray}
where the following diagrammatic notations are used:
$
 \raisebox{-.13cm}{\epsfysize .4cm\epsfbox{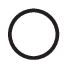}}
 \,-\, S^z,
$
$
 \raisebox{-.13cm}{\epsfysize .4cm\epsfbox{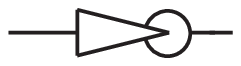}}
$
$
 \,-\, g_i(\omega_n),
$
wavy line is the Fourier transform of the hopping integral $t_{\bk }$.
 Semi-invariants are represented by ovals and contain the
$\delta$-symbols on site indices.
  In the spirit of the traditional mean field approach, renormalization of
the basic semi-invariant by insertion of independent
loop fragments is taken into account in (\ref{Sz_diagr}).
 The analytical expression for the loop is the following:
\begin{eqnarray}
 \label{loop1}
\fl
 \raisebox{-.4cm}{\epsfysize 1.1cm\epsfbox{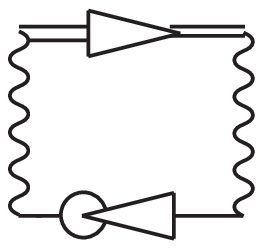}}
 =\frac 2N \sum_{n,\bk}\frac{t^2_{\bk}}{g^{-1}(\omega_n)-
 t_{\bk}}\left(\frac{P^+_i}{\rmi\omega_n-\varepsilon_1}+
 \frac{P^-_i}{\rmi\omega_n-\varepsilon_2}\right)
 =\beta(\alpha_1P^+_i+\alpha_2P^-_i).
\end{eqnarray}

 It should be noted that within the presented self-consistent scheme
of the GRPA, chain fragments form the single-electron Green's
function in the Hubbard-I approximation, and in the sequences of
loop diagrams in expressions for thermodynamic and pair correlation
functions the connections between any two loops by more than
one semi-invariant are omitted.
 This procedure includes renormalization of the higher order
semi-invariants similar to the one given by expression (\ref{Sz_diagr}).

 Summation of diagrammatic series (\ref{Sz_diagr}) for the pseudospin mean value
is equivalent to averaging with the mean-field type Hamiltonian:
\begin{eqnarray}
\label{MF}
 H_{\rm MF}{=}\sum\limits_i\{\varepsilon_1
 (n_{i\uparrow}{+}n_{i\downarrow})+
 \varepsilon_2(\tilde{n}_{i\uparrow}{+}\tilde{n}_{i\downarrow})-
 yS^z_i\},
\end{eqnarray}
where an expression $\quad y=h+\alpha_2-\alpha_1 \quad $ determines an
internal effective self-consistent field formed by retarded
many-body interaction between pseudospins via conducting electrons.

 Equation for pseudospin mean value in the uniform case
is as follows
\begin{eqnarray}
 \label{b_y1}
 \langle S^z_l\rangle=\langle S^z\rangle_{\rm MF}=b(y),
\end{eqnarray}
where
\begin{eqnarray}
 \label{b_y2}
 b(y)=\frac 12 \tanh\left\{\frac{\beta}{2}y+
 \ln{\frac{1+\rme^{-\beta\varepsilon_1}}
 {1+\rme^{-\beta\varepsilon_2}}}
\right\} .
\end{eqnarray}
 An analytical expression for mean value of the particle number
can be presented in the form \cite{Tabunshchyk1}:
\begin{eqnarray}
\label{concentration}
\fl
\langle n \rangle=\frac 2N
\sum_{\bk}\Big[n(\varepsilon_{\rs I}(t_{\bk}))
+n(\varepsilon_{\rs II}(t_{\bk}))\Big]
-[1+2\langle S^z\rangle]
n(\varepsilon_2)-[1-2\langle S^z\rangle]
n(\varepsilon_1).
\end{eqnarray}
 Here $n(\varepsilon)=\frac{\displaystyle 1}
{\displaystyle 1+\rme^{\beta\varepsilon}}$
is Fermi distribution.

 Diagrammatic equation on pseudospin correlator $\langle
S^zS^z\rangle_{\bq}$ within the framework of the GRPA
has the following form:
\begin{eqnarray}
\label{szsz}
\fl
\langle S^zS^z\rangle_{\bq} =
 \raisebox{-.6cm}[.45cm][.7cm]{\epsfysize 1.3cm\epsfbox{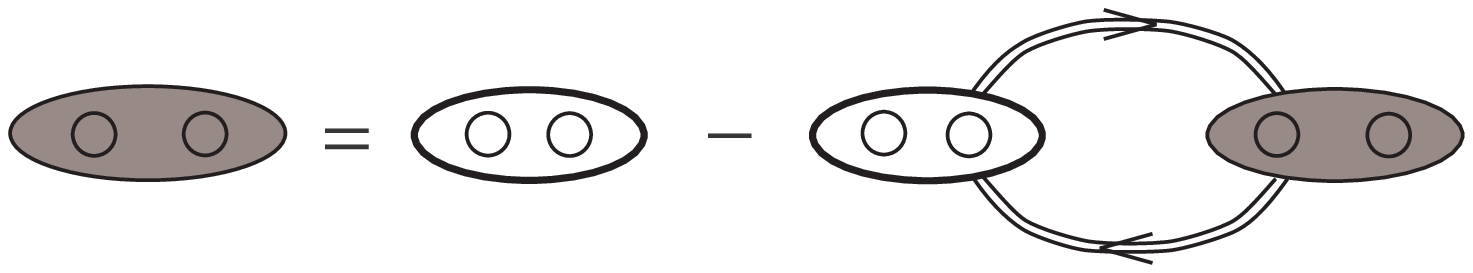}}\quad .
\end{eqnarray}
 Equation (\ref{szsz}) differs from the corresponding one for the Ising model
in RPA by the replacement of the direct exchange interaction by the
electron loop
$
 \Pi_{\bq}=
 \raisebox{-0.4cm}{ \epsfysize .85cm\epsfbox{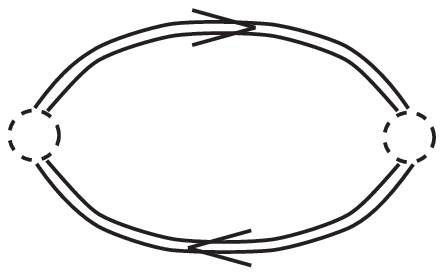}}\;
$
(describing the many-body retarded interaction
between pseudospins via conducting electrons)
\begin{eqnarray}
 \label{petlia}
\fl \Pi_{\bq}=\frac 2N \sum\limits_{n,\bk}
    \Lambda^2_n\tilde{t}_n(\bk)\tilde{t}_n(\bk{+}\bq),\;\;
 \Lambda_n=\frac g
 {(\rmi\omega_n {+}\mu)^2{-}g^2/4},
 \;\; \tilde{t}_n(\bk)=\frac{t_{\bk}}{(1-g_nt_{\bk})}.
\end{eqnarray}
 The first term in equation (\ref{szsz}) is the zero-order correlator
renormalized due to the inclusion in all basic semi-invariants of
the mean-field type contributions (`single-tail' parts)
coming from the effective pseudospin interaction
\begin{eqnarray}
\label{correlator}
\fl \raisebox{-.1cm}[1.2cm][.cm]{\epsfxsize 11cm\epsfbox{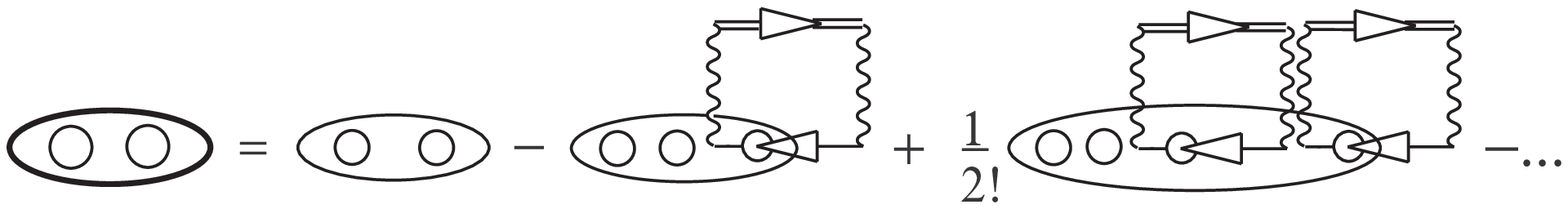}}
\end{eqnarray}
and is thus calculated with the help of Hamiltonian $H_{\rm MF}$.

 In analytical form, solution of equation (\ref{szsz}) is
equal to
\begin{eqnarray}
\fl
 \langle S^zS^z\rangle_{\bq}=
 \frac{1/4 - \langle S^z\rangle^2}
     {1+\frac 2N \sum\limits_{n,\bk}
 \Lambda^2_n\tilde{t}_n(\bk)\tilde{t}_n(\bk+\bq)
 (1/4 - \langle S^z\rangle^2)},
\end{eqnarray}
and is different from zero only in a static case ($\omega_n=0$)
(since pseudospin operator commutes with the initial
Hamiltonian).

 Within the same approach the grand canonical potential was
obtained \cite{Tabunshchyk1}.
 The corresponding analytical expression is
\begin{eqnarray}
\fl\Omega_{\rm GRPA}={-}\frac 2{N\beta}\sum\limits_{n,\bk}\ln(1{-}t_{\bk}g(\omega_n))
       -\frac 2{N\beta}\sum\limits_{n,\bk}\frac{g(\omega_n))t_{\bk}^2}
                                               {g^{-1}(\omega_n))-t_{\bk}}
       -\frac 1\beta \ln {\Tr}(\rme^{-\beta H_{\rm MF}} ).
\end{eqnarray}
 It was shown that all quantities can be derived from the grand canonical
potential
\[
\fl
 \frac{\rmd\Omega}{\rmd (-\mu)}=\langle n \rangle,\quad
 \frac{\rmd\Omega}{\rmd (-h)}=\langle S^z\rangle,\quad
 \frac{\rmd\langle S^z\rangle}{\rmd (\beta h)}
 =\langle S^zS^z\rangle_{{\bq}=0},
\]
what demonstrates thermodynamic consistence of the proposed
approximation.

 In the $\mu =const$ regime the stable states are determined from
the minimum of the grand canonical potential.
 The solution of equation for the pseudospin mean value and
calculation of potential $\Omega$ (as well as the pair
correlation function) were performed numerically
\cite{Tabunshchyk1}.
 The first order phase transitions between the different uniform phases
(bistability effect) at change of temperature $T$
and field $h$
can take place
when chemical potential is placed within the electron subbands.
 In the case when chemical potential is fixed and placed between the
electron subbands, the uniform phase become unstable with respect to fluctuations
with $\bq=(\pi,\pi)$, and the possibility of second order phase transitions
between the uniform and chess-board phase exists at the change
of temperature or field \cite{Tabunshchyk3}.

\begin{figure}[ht]
\begin{center}
 \epsfxsize 5.cm\epsfbox{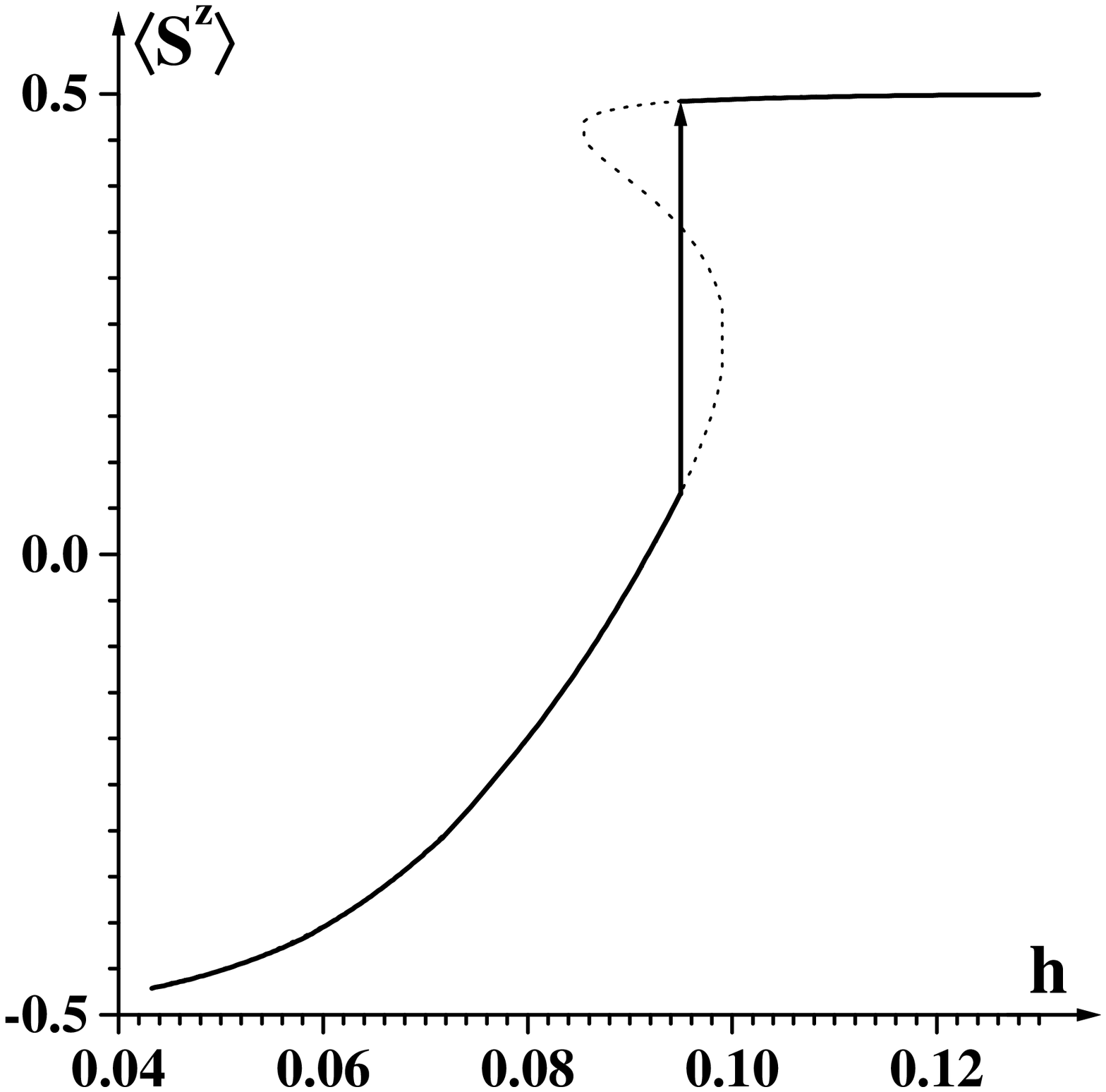}\qquad
 \epsfxsize 5.cm\epsfbox{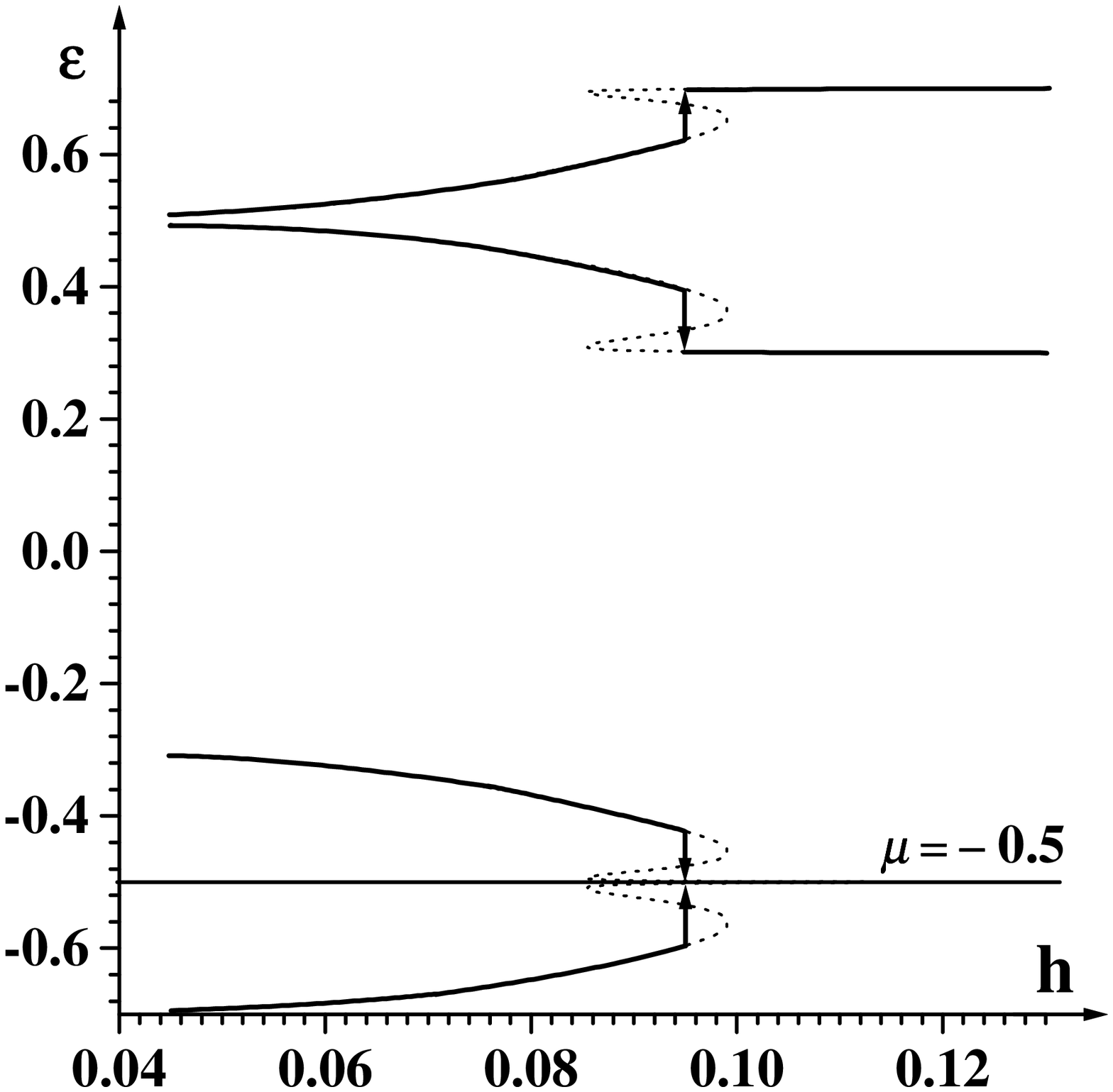}
\end{center}
\caption{Field dependences of pseudospin mean value and
         electron bands boundaries for $\mu=const$ regime when
         chemical potential is placed in the lower subband
         (in the self-consistent GRPA)
         ($g=1$, $t_{\bk=0}=0.2$, $\mu=-0.5$, $T=0.0132$)}
 \label{Fig1}
\end{figure}

 As mentioned above, the band structure is determined by the pseudospin
mean value (figure~\ref{Fig1}), and its change is accompanied by the
corresponding changes of the electron concentration
(see \cite{Tabunshchyk1,Tabunshchyk2} for details).

\begin{figure}[ht]
\begin{center}
 \raisebox{-2.5cm}[2.5cm][2.5cm]
 {\epsfxsize 5.5cm\epsfbox{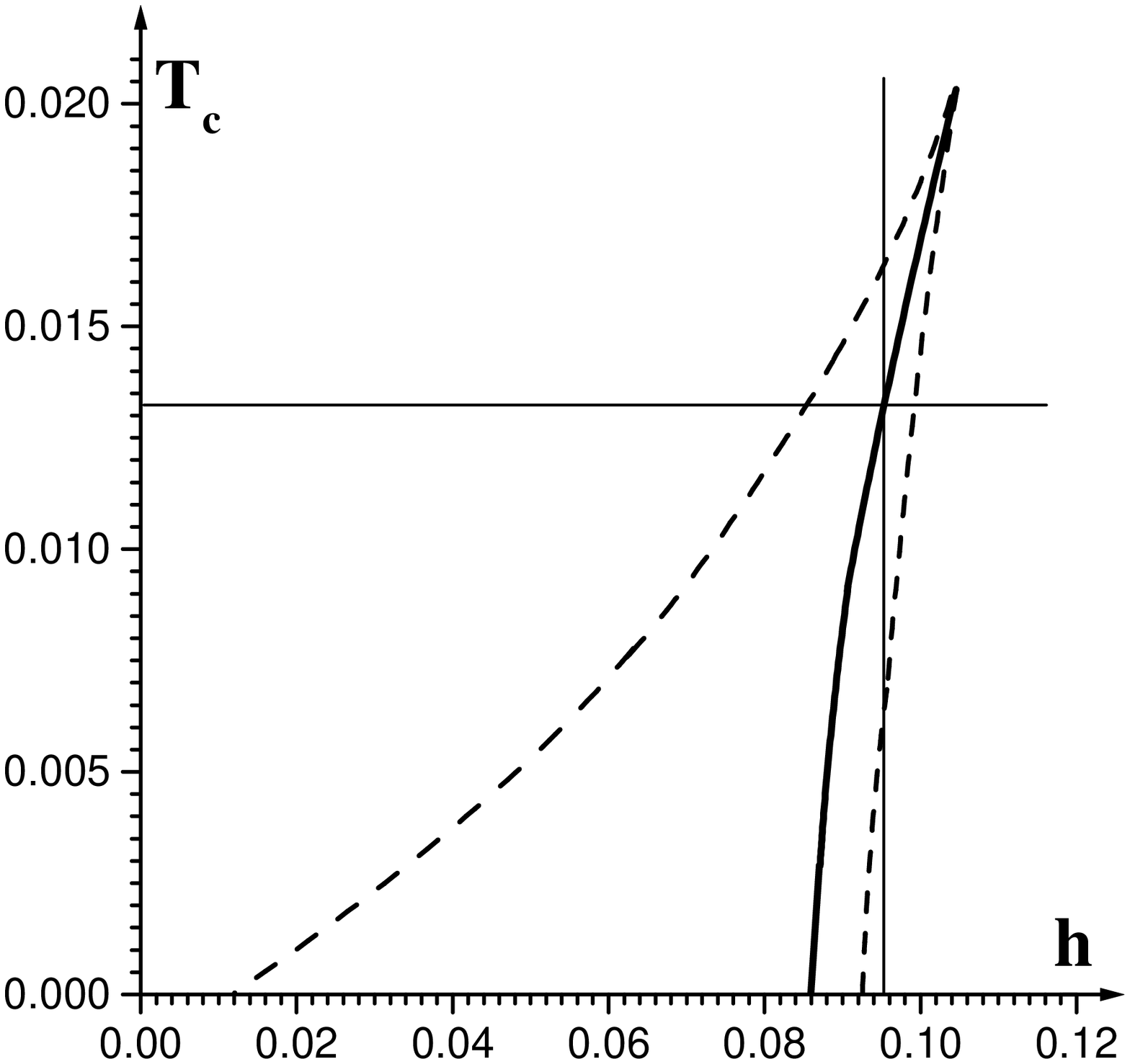}}\; (a)\qquad
 \raisebox{-2.5cm}[2.5cm][2.5cm]
 {\epsfxsize 5.6cm\epsfbox{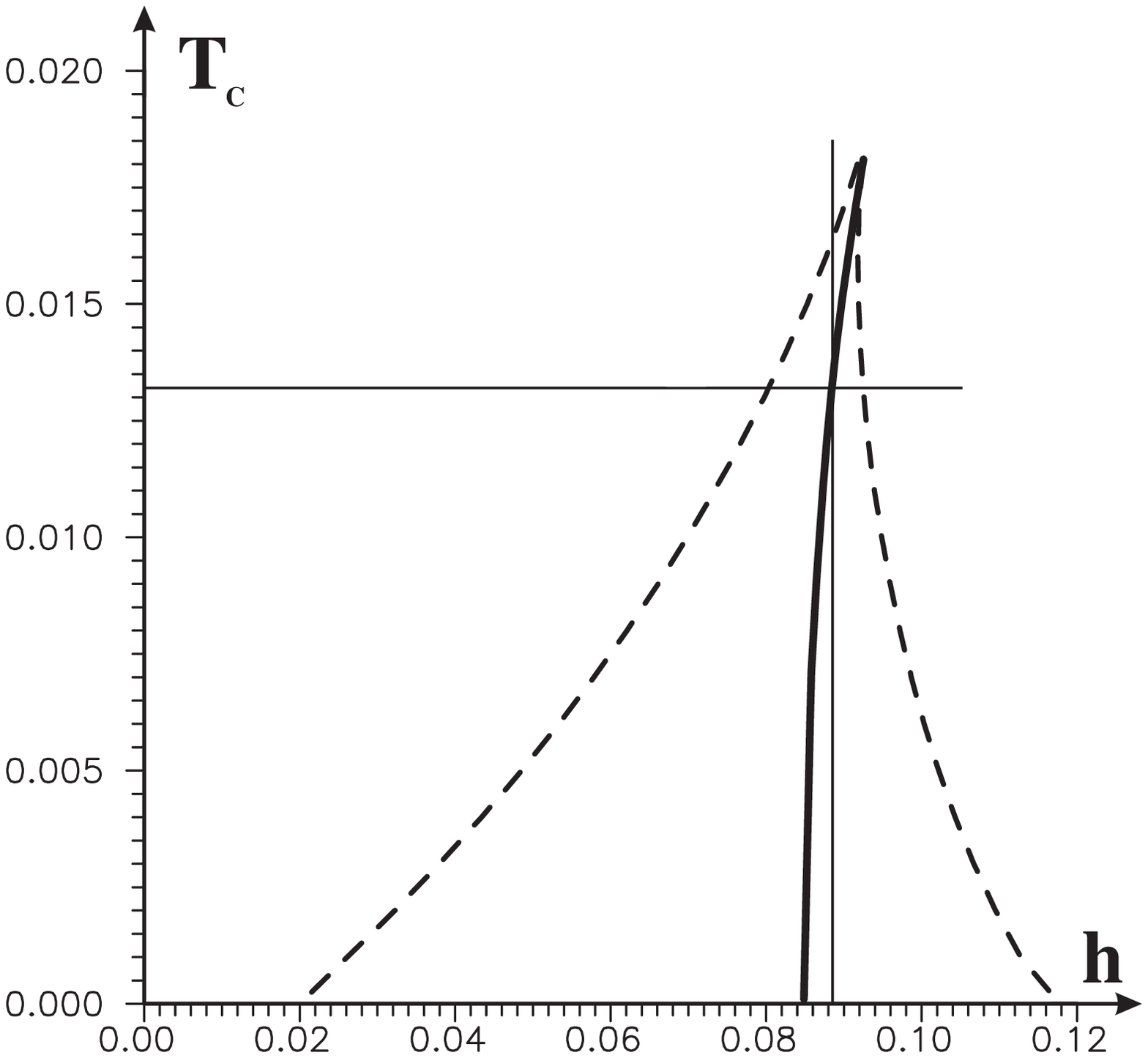}}\; (b)
\end{center}
\caption{$T_{\rm c}-h$ phase diagrams:
        (a) within the self-consistent GRPA,
        (b) within the DMFA \cite{Shvaika}.
        Solid and dashed lines indicate the first order phase transition
        line and boundaries of the phase stability (spinodal lines),
        respectively ($g=1$, $t_{\bk=0}=0.2$, $\mu=-0.5$)}
 \label{Fig2}
\end{figure}

 The simplified version of PEM was considered in \cite{Shvaika} within
the framework of DMFA scheme as well.
 The obtained phase diagrams within the self-consistent GRPA
and DMFA  are presented in figure~\ref{Fig2}
(in the case when chemical potential
is placed in the center of lower subband $\mu=-0.5$).

 One can see a quite sufficient coincidence of shapes of the first order
phase transition curves (thick solid line).
 But unlike to the phase diagram in figure~\ref{Fig2}b obtained within the DMFA,
boundaries of the phase stability region calculated in the self-consistent GRPA
have the same type of slope, hence a region exists, where the vertical line twice
crosses the boundary of the phase stability region (figure~\ref{Fig2}a).
 The analysis of the $\langle S^zS^z\rangle$ behaviour in this
region (for fixed value of the longitudinal field $h=0.0875$)
with the decrease of temperature shows that high temperature phase
becomes unstable with respect to fluctuations with $\bq = 0$
(the lower crossing point of the vertical line and
boundary of the phase stability region in figure~\ref{Fig2}a.)
 Similar results were obtained previously in \cite{Stasyuk2,Stasyuk3}
for temperature behaviour of the correlation
functions in the case of infinite single-site electron correlation
$U\rightarrow\infty$ within the framework of the GRPA.

 In the case of the fixed value of the electron concentration
condition of equilibrium is determined by the minimum of free
energy.
 In this regime the first
order phase transition with a jump of the pseudospin mean value
(accompanied by the change of electron concentration) transforms
into a phase separation into the regions with different electron
concentrations and pseudospin mean values \cite{Tabunshchyk1}.
 For the first time the instability with respect to the phase separation in
pseudospin-electron model was marked in \cite{Stasyuk1}, where
it was obtained within the GRPA in the limit $U\rightarrow\infty$.

\section{Self-consistent gaussian fluctuation approximation}
 The above considered approach takes into account only
contributions of mean field type.
 In this section we present the developed in \cite{Tabunshchyk}
consistent scheme for calculation of thermodynamic and
correlation functions, which takes into account the
gaussian fluctuations of the self-consistent
mean field.
 We also reduce the presented here method to more suitable for numerical
calculation taking into account a restricted class of diagrams.
 Such an approximation for spin models with a direct interaction
proposed by Onyszkiewicz \cite{Onyshkevych1,Onyshkevych2}
yields a much better description of critical properties of spin
models in the whole range of temperature then others.

 In constructing a higher order approximation that takes into account
the fluctuation effects of the self-consistent mean field,
we use the self-consistent GRPA as the zero-order one.
 This means that all `single-tail' parts of diagrams (\ref{Sz_diagr})
and (\ref{correlator}) are already summed up and all semi-invariants
are calculated using the distribution with the Hamiltonian
$H_{\rm MF}$ (\ref{MF}).
 All these semi-invariants are represented graphically by thick ovals
and contain the $\delta$-symbols on site index:
\begin{eqnarray}
\label{semiinv}
\fl\raisebox{-.09cm}{\epsfysize .4cm\epsfbox{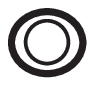}} {=}
\langle S^z\rangle_{\rm MF} = b(y),\;
\raisebox{-.09cm}{\epsfysize .4cm\epsfbox{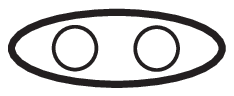}} {=}
\langle S^zS^z\rangle^{\rm c}_{\rm MF} = \frac{\partial b(y)}
{\partial\beta y},\;
\raisebox{-.09cm}{\epsfysize .4cm\epsfbox{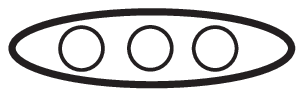}} {=}
\langle S^zS^zS^z\rangle^{\rm c}_{\rm MF} =
\frac{\partial^2 b(y)}{\partial(\beta y)^2}.
\end{eqnarray}

 As an approximation that goes beyond of the self-consistent GRPA we use,
similarly to \cite{Izyumov3,Izyumov2}, the approach taking into
account the so-called `double-tail' diagrams.
 A corresponding diagrammatic series for the pseudospin mean value
can be written as
\begin{eqnarray}
 \label{sz_gaus}
 \fl
 \langle S^z\rangle=
 \raisebox{-.25cm}[1.4cm][.1cm]{\epsfxsize 9.cm\epsfbox{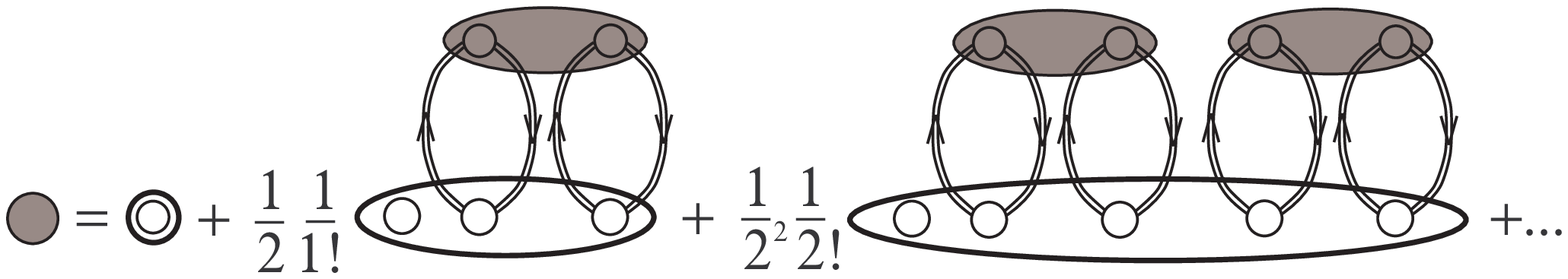}}\quad .
\end{eqnarray}
 The diagrammatic equation for the pseudospin correlator $\langle S^zS^z\rangle_{\bq}$
within the presented here approximation is similar to the corresponding one
within the GRPA (\ref{szsz}) and given by
\begin{eqnarray}
\label{sz_sz}
 \fl
 \langle S^zS^z\rangle_{\bq} =
 \raisebox{-.65cm}[.75cm][.75cm]{\epsfysize 1.4cm\epsfbox{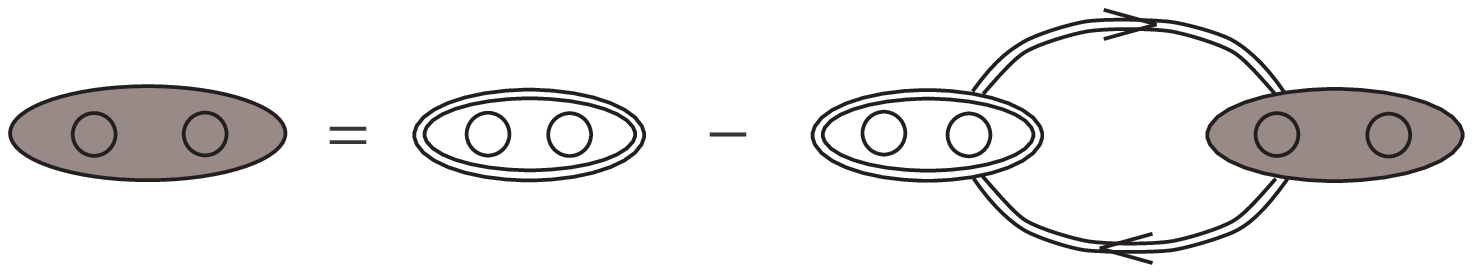}}\quad ,
\end{eqnarray}
but now all semi-invariants in this equation are renormalized due to the
`double-tail' parts.
 Thus the corresponding diagrammatic series
on zero-order correlator looks like
\begin{eqnarray}
 \label{szsz_g}
 \fl
 \Xi = \raisebox{-.26cm}[1.2cm][.1cm]{\epsfxsize 11.cm\epsfbox{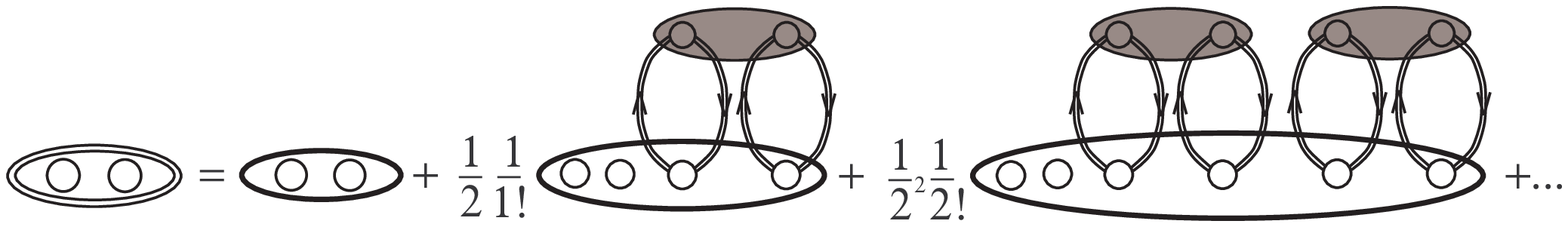}}\quad .
\end{eqnarray}
 The contribution corresponding to the `double-tail' fragment
of the diagrams can be written in the following analytical form
(using the notation (\ref{petlia})):
\begin{eqnarray}
\label{X}
\fl
X=\raisebox{-.6cm}{\epsfysize 1.2cm\epsfbox{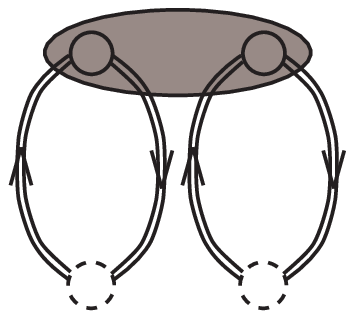}}
=\frac {2^2}{N^3}\sum\limits_{n,n'}
\sum\limits_{\bk,\bk'}\sum\limits_{\bq}
\Lambda^2_n\tilde{t}_n(\bk)\tilde{t}_n(\bk{-}\bq)
\langle S^zS^z\rangle_{\bq}
\Lambda^2_{n'}\tilde{t}_{n'}(\bk')\tilde{t}_{n'}(\bk'{+}\bq).
\end{eqnarray}
 The equation on pseudospin correlation function
(\ref{sz_sz}) has such a solution:
\begin{eqnarray}
\label{correlatior}
\langle S^zS^z\rangle_{\bq} =
\frac \Xi {1+\frac 2N \sum\limits_{n,\bk}
  \Lambda^2_n\tilde{t}_n(\bk)\tilde{t}_n(\bk{+}\bq)\Xi}.
\end{eqnarray}
 Since the pseudospin correlator (\ref{correlatior}) is a frequency
independent function, in the expression for $X$ (\ref{X}) we have two independent
sums over internal Matsubara frequencies, allowing one (by decomposition
into simple fractions) to sum over all internal frequencies.

 The diagrammatic series (\ref{sz_gaus}) and (\ref{szsz_g}) can be
expressed in the following analytical forms \cite{Tabunshchyk}:
\begin{eqnarray}
\label{sz_g}
 \langle S^z\rangle
  = \frac 1{\sqrt{2\pi X}}\int\limits^{+\infty}_{-\infty}
 \exp\Big({-}\frac {\xi^2}{2X}\Big)b(y+\xi)\rmd\xi,
\end{eqnarray}
\begin{eqnarray}
\label{sz_g1}
 \Xi
 = \frac 1{X\sqrt{2\pi X}}\int\limits^{+\infty}_{-\infty}
 \exp\Big({-}\frac {\xi^2}{2X}\Big)\xi b(y+\xi)\rmd\xi.
\end{eqnarray}
 As one can see the contribution of diagrammatic series with `double-tail' parts
corresponds to the average with the Gaussian distribution, where
$X$ can be interpreted as the root--mean--squares (r.m.s.) fluctuation
of the mean field around the mean value of $y$.
 Thus we obtain a self-consistent set of equations
for pseudospin mean value (\ref{sz_g}) and r.m.s. fluctuation
parameter (\ref{X}) as well as
expression for pseudospin correlation function (\ref{correlatior}).

  The grand canonical potential within  the approximation
presented here is given by the diagrammatic series below:
\begin{eqnarray}
\label{omega}
\fl
\beta\Omega=\beta\Omega_{\rm GRPA}\!\!\!
 \raisebox{-1.85cm}{\epsfxsize 8.3cm\epsfbox{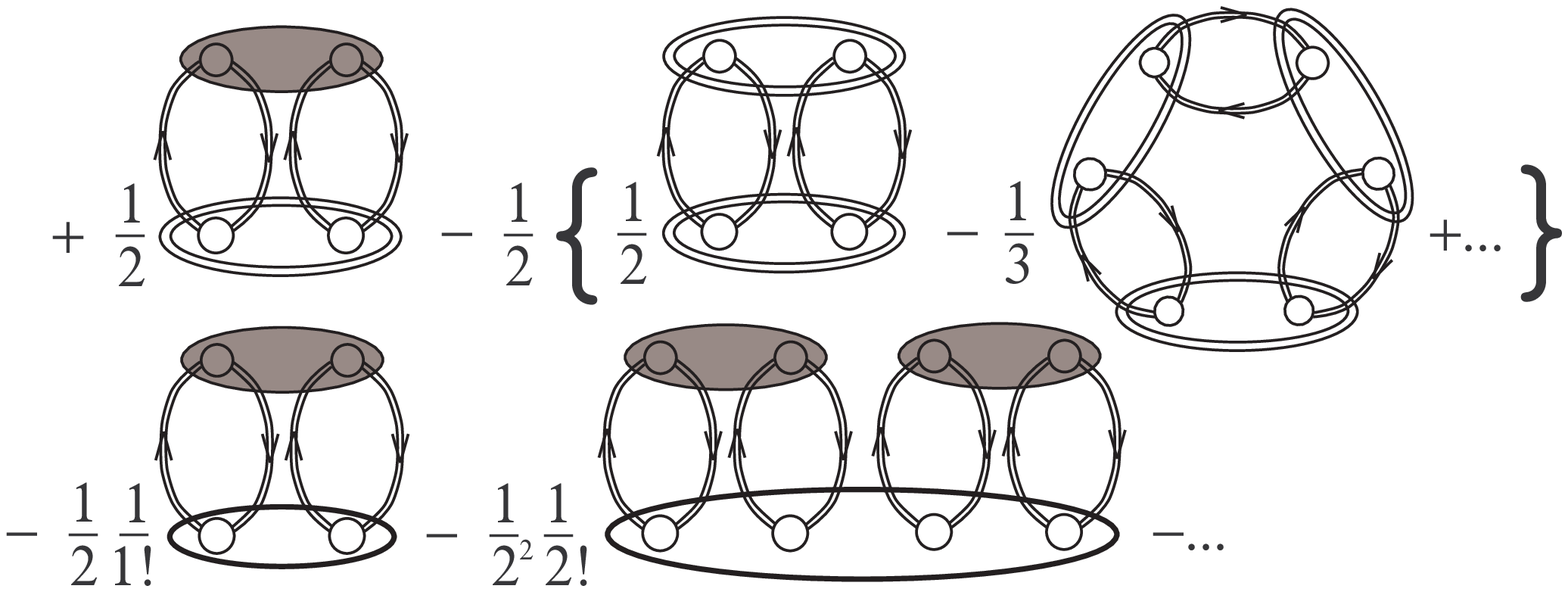}}
\end{eqnarray}
 The corresponding analytical expression is
\begin{eqnarray}
 \nonumber
\fl
\Omega=\Omega_{\rm GRPA}{+}\frac 12
\frac 1{\sqrt{2\pi X}}\int\limits^{+\infty}_{-\infty}
 \rme^{{-}\frac {\xi^2}{2X}}\xi b(y{+}\xi)\rmd\xi
 -\frac 12\int\limits^{+\infty}_{-\infty}
\Big \{1{-}{\rm erf}\bigg(\frac {|\xi|}{\sqrt{2X}}\bigg)\Big\}{\rm sign}(\xi)
 b(y{+}\xi)\rmd\xi\\
\lo  -\frac 12\Xi\frac 2N \sum\limits_{n,\bk}
      \Lambda^2_n\tilde{t}_n(\bk)^2
       +\frac 12\ln \bigg(1+\Xi\frac 2N \sum\limits_{n,\bk}
      \Lambda^2_n\tilde{t}_n(\bk)^2\bigg).
\end{eqnarray}

 The grand canonical potential written
in this form satisfies the stationary conditions:
\begin{eqnarray}
\label{condition1}
\frac{\rmd \Omega}{\rmd X}=0,\quad
\frac{\rmd\Omega}{\rmd\langle S^z\rangle}=0,
\end{eqnarray}
which are equivalent to the equations (\ref{X}) and (\ref{sz_g}).
 The consistency of the approximations used for the pseudospin mean value, pseudospin
correlation function and grand canonical potential can be checked
explicitly using the relations:
\begin{eqnarray}
\label{condition2}
\frac{\rmd\Omega}{\rmd (-h)}=\langle S^z\rangle,\quad
\frac{\rmd\langle S^z\rangle}{\rmd (-h\beta)}\Big |_{X={\rm const}}=
\langle S^zS^z\rangle_{\bq = 0}.\quad
\end{eqnarray}
 In the limit of vanishing fluctuations our results go
over into the ones obtained within the self-consistent GRPA.

 The analytical scheme presented here for the pseudospin-electron
model can be easily reduced to the scheme proposed by Onyszkiewicz
for spin models.
 For this purpose we consider renormalization with use
of the simplest possible pseudospin correlation function
involving only gaussian fluctuations of the mean field
for the `double-tail' fragment of the diagram
\begin{eqnarray}
 \fl \epsfxsize 10.cm\epsfbox{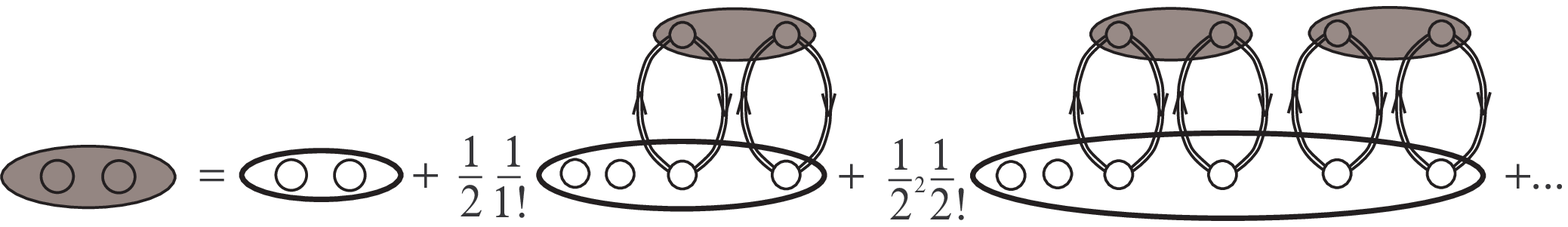}
\end{eqnarray}
 Within the framework of this simplification the grand canonical
potential satisfies the stationary conditions (\ref{condition1})
and can be written in the following analytical form:
\begin{eqnarray}
\fl
\Omega = \Omega_{\rm GRPA}+\frac 14 \Xi X
 -\frac 12\int\limits^{+\infty}_{-\infty}
\Big \{1-{\rm erf}\bigg(\frac {|\xi|}{\sqrt{2X}}\bigg)\Big\}{\rm sign}(\xi)
 b(y+\xi)\rmd\xi.
\end{eqnarray}
 The diagrammatic series for pseudospin mean value is the same
as the above presented ones (\ref{sz_gaus}).
 Finally, the set of equations on pseudospin mean value and r.m.s.
fluctuation parameter can be written down as:
\begin{eqnarray}
\label{Sz_on}
\fl \langle S^z\rangle=
 \frac 1{\sqrt{2\pi X}}\int\limits^{+\infty}_{-\infty}
 \exp\Big({-}\frac {\xi^2}{2X}\Big)b(y+\xi)\rmd\xi,
\end{eqnarray}
\begin{eqnarray}
\label{X_on}
\fl X&=&\left(\frac 2N\sum\limits_{\bk,n}
 \Lambda^2_n{\tilde{t}\,}^2_n(\bk)\right)^2
 \frac 1{X\sqrt{2\pi X}}\int\limits^{+\infty}_{-\infty}
 \exp\Big({-}\frac {\xi^2}{2X}\Big)\xi b(y+\xi)\rmd\xi.
\end{eqnarray}

\section{Phase diagrams within the gaussian fluctuation approach}
 Solution of equations for the pseudospin mean value and r.m.s. fluctuation
parameter is performed numerically for the square lattice with
nearest-neighbor hopping.
 The stable state of the system is described by the solution
(from the possible set of ones)
corresponding to the global minimum of grand canonical potential;
metastable states are related to solutions corresponding to
local minima.
\begin{figure}[htb]
\begin{center}
 \raisebox{-2.5cm}[2.9cm][2.4cm]
 {\epsfxsize 5.5cm\epsfbox{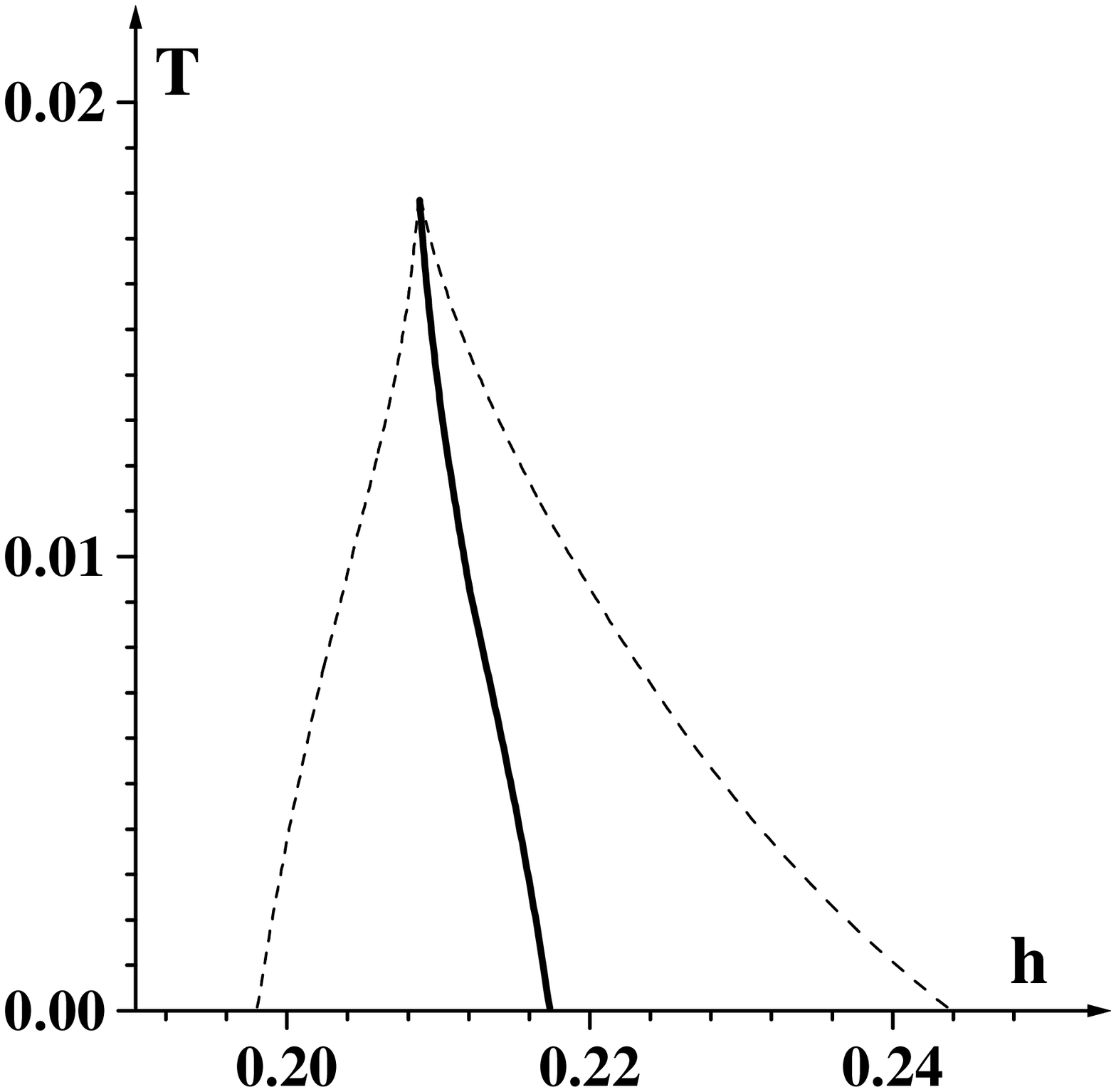}}\; (a)\qquad
 \raisebox{-2.5cm}[2.9cm][2.4cm]
 {\epsfxsize 5.5cm\epsfbox{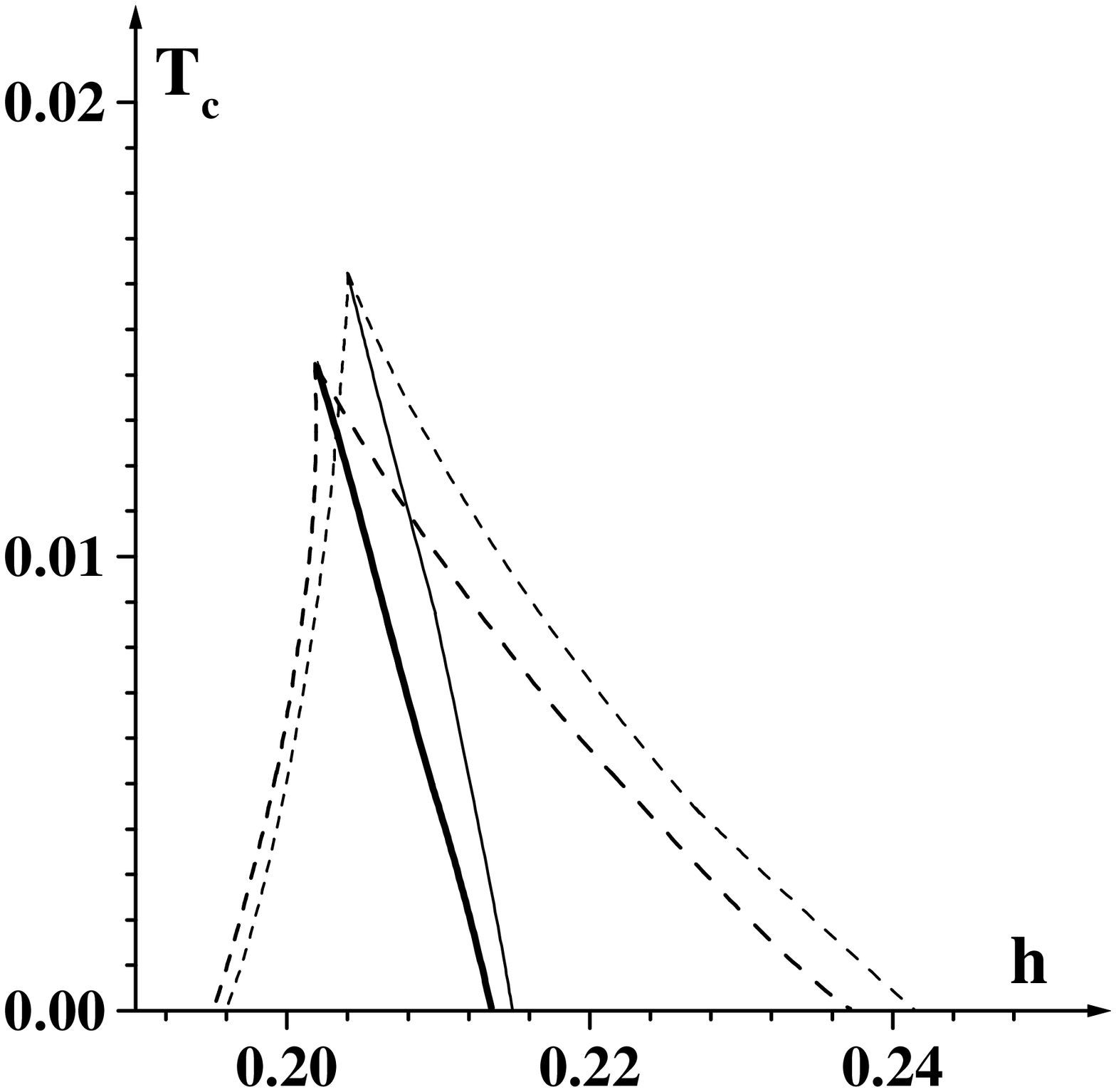}}\; (b)
\end{center}
\caption{$T_{\rm c}-h$ phase diagrams:
        (a) within the self-consistent GRPA,
        (b) within the gaussian fluctuation methods.
        ($g=1$, $t_{\bk=0}=0.2$, $\mu=-0.4$).}
 \label{Fig3}
\end{figure}

 We would like to note that there is no particular difference (figure~\ref{Fig3}b)
between the results obtained within the self-consistent gaussian
fluctuation approach by solving the set of equations (\ref{X}), (\ref{sz_g})
and its simplified version (the set of equations (\ref{Sz_on}), (\ref{X_on})):
the same topological type and slope of phase diagrams,
similar field and temperature behaviours of pseudospin mean
value and grand canonical potential were observed.
 Only small quantitative differences (as one can see in phase
diagrams in figure~\ref{Fig3}b) are seen (the thin solid phase coexistence
line is obtained in the self-consistent gaussian fluctuation approach,
thick solid line corresponds to the Onyszkiewicz type
approximation).
 Therefore, to show the influence of the gaussian fluctuations
on the thermodynamic properties of the PEM we perform all calculations
for the simplified variant of the gaussian fluctuation
approach; their results are presented below.

 In the case when chemical potential is fixed and placed within the electron
subbands (as one can see comparing figures~\ref{Fig3}a and~\ref{Fig3}b for $\mu=-0.4$)
taking into account of fluctuations does not change qualitatively the
results obtained previously within the self-consistent GRPA \cite{Tabunshchyk1}
(except for some special cases presented below).
 The quantitative changes due to fluctuations are important in the region of
the critical point leading to significant lowering of the critical point temperature
(T=0.018 for the self-consistent GRPA in figure~\ref{Fig3}a and T=0.0145 for the
Onyszkiewicz type approximation in figure~\ref{Fig3}b).

\begin{figure}[htb]
\begin{center}
 \raisebox{-2.3cm}[2.4cm][2.4cm]
 {\epsfxsize 5.cm\epsfbox{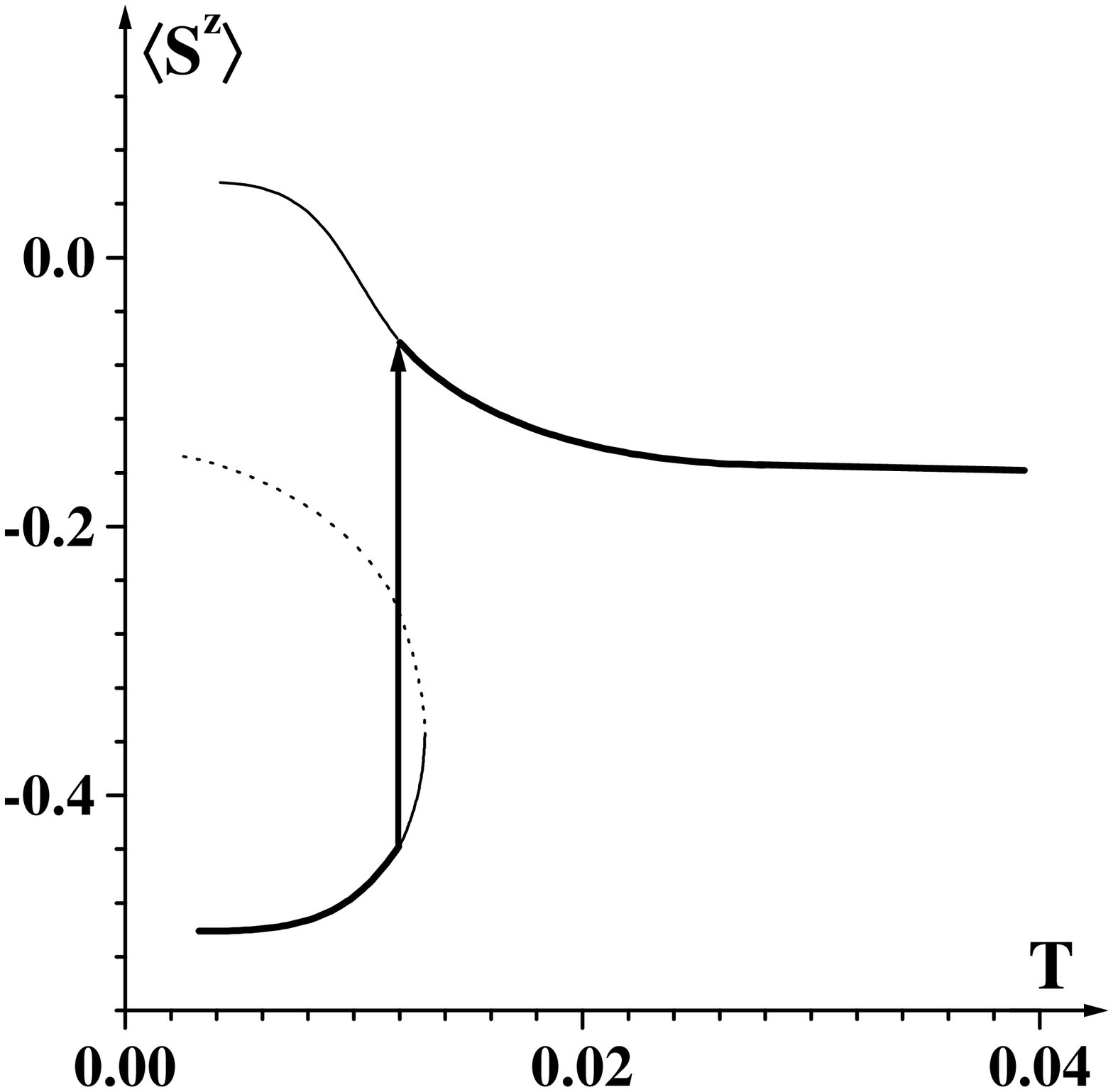}}\; (a)\qquad
 \raisebox{-2.3cm}[2.4cm][2.4cm]
 {\epsfxsize 5.cm\epsfbox{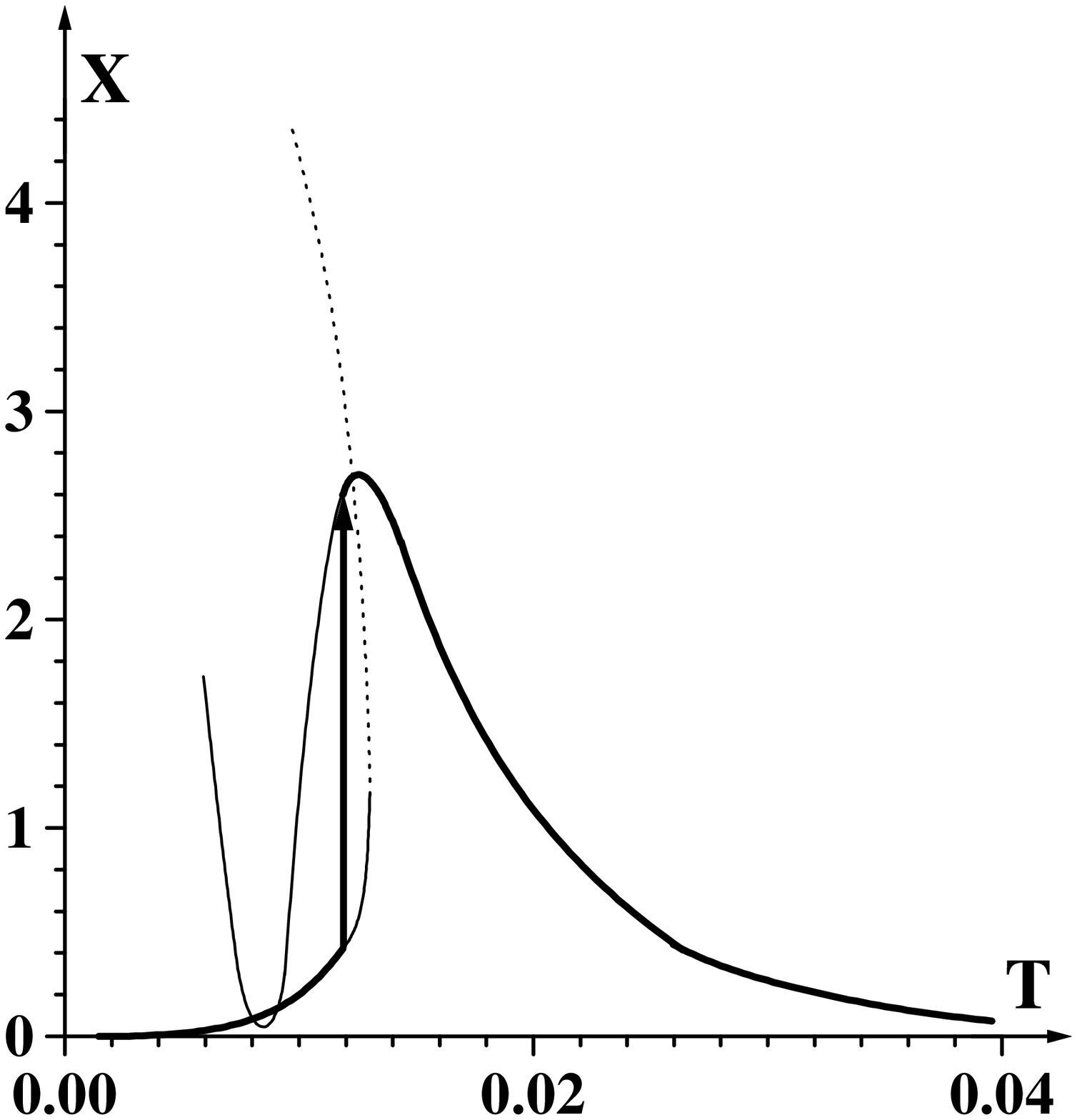}}\; (b)
\end{center}
\caption{Temperature dependences of
        (a) pseudospin mean value and
        (b) r.m.s. fluctuation parameter
        ($g=1$, $t_{\bk=0}=0.2$, $\mu=-0.4$, $h=0.204$).
         Thick solid lines correspond to thermodynamically stable states,
         other lines denote metastable and unstable ones.}
 \label{Fig4}
\end{figure}

 Temperature behaviour of the pseudospin mean value and r.m.s.
fluctuation parameter is presented in figure~\ref{Fig4} for the fixed value of the
longitudinal field $h=0.204$.
 At increasing temperature, the pseudospin mean value
and r.m.s. parameter jump from the branch of the low
temperature phase to the one of the high temperature phase
in the phase transition point $T=0.0109$.

\begin{figure}[htb]
\begin{center}
 \epsfxsize 5.5cm\epsfbox{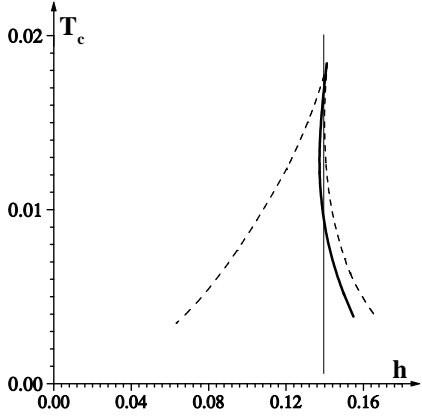}
\end{center}
 \caption{$T_{\rm c}-h$ phase diagrams
         ($g=1$, $t_{\bk=0}=0.2$, $\mu=-0.5$).}
 \label{Fig5}
\end{figure}

 A certain qualitative change takes place (figure~\ref{Fig5}) when the chemical
potential is placed near the center of electron subband ($\mu=-g/2=-0.5$).
 Comparing to the self-consistent GRPA (figure~\ref{Fig2}a)
or the DMFA (figure~\ref{Fig2}b),
a change of slope of the phase coexistence line is observed.
 The vertical line on the $T_{\rm c} - h$ phase diagram (figure~\ref{Fig5}) twice crosses
the phase coexistence curve and hence there exists the possibility of the two
sequential first order phase transitions (reentrance) with change of temperature.

\begin{figure}[htb]
\begin{center}
 \raisebox{-2.5cm}[2.95cm][2.4cm]
 {\epsfxsize 5.5cm\epsfbox{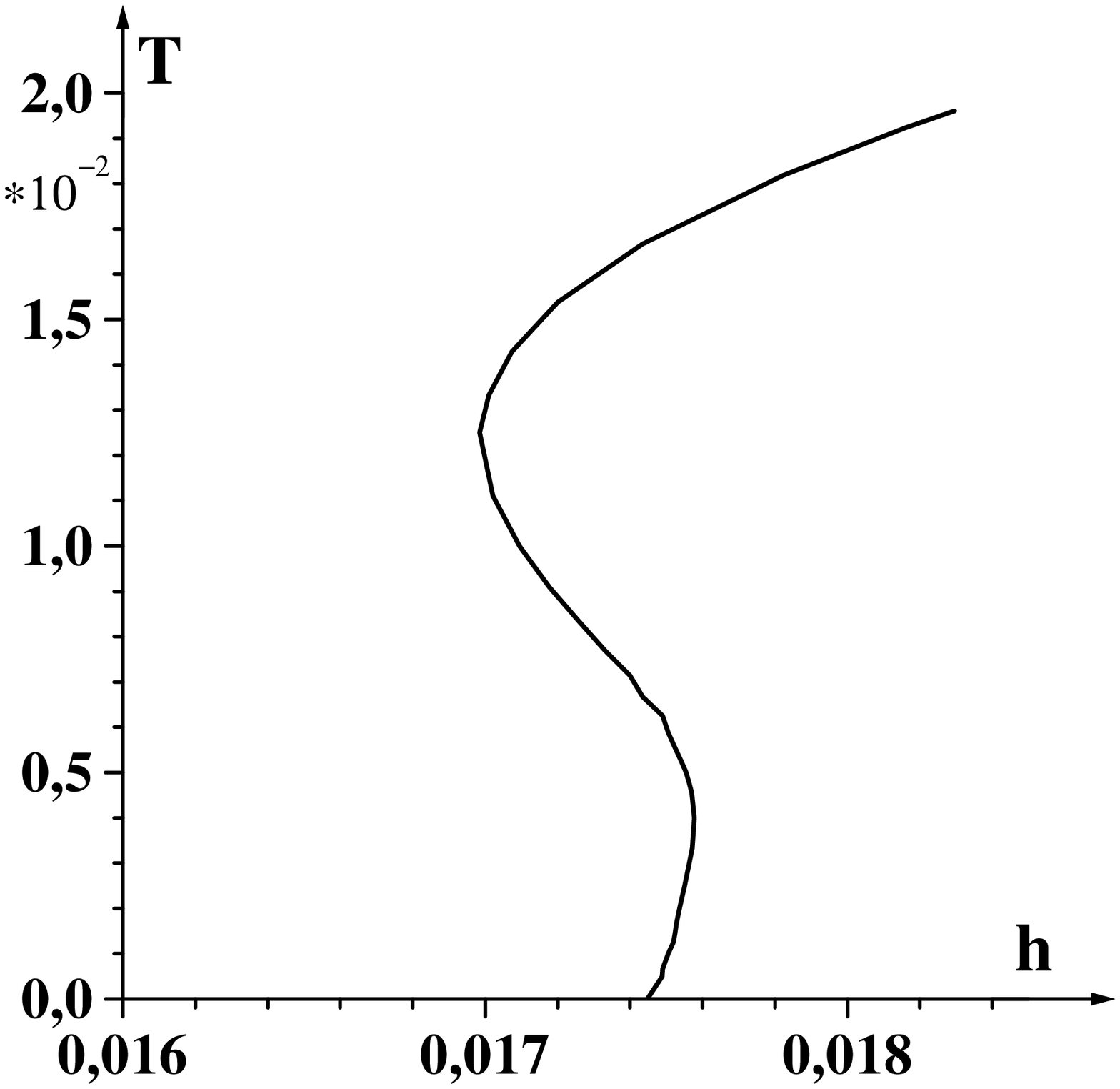}}\; (a)\qquad
 \raisebox{-2.5cm}[2.9cm][2.4cm]
 {\epsfxsize 5.5cm\epsfbox{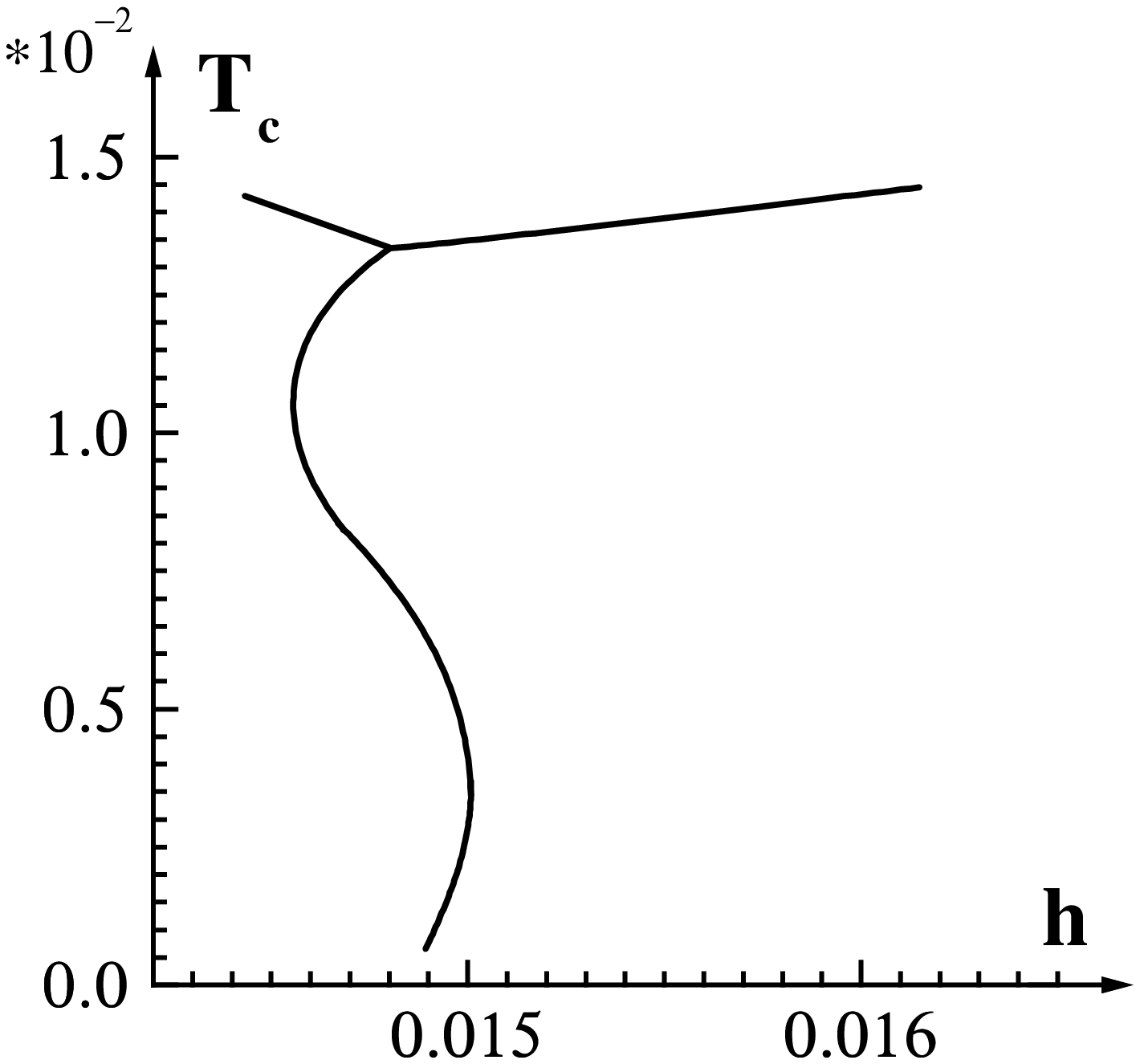}}\; (b)
\end{center}
\caption{$T_{\rm c}-h$ phase diagrams:
        (a) self-consistent GRPA,
        (b) gaussian fluctuation method
        ($g=1$, $t_{\bk=0}=0.2$, $\mu=-0.6$).}
 \label{Fig6}
\end{figure}

 For the different values of chemical potential, a slope of the
phase coexistence curve can vary
($\mu=-0.5$ in figure~\ref{Fig2} and $\mu=-0.4$ in figure~\ref{Fig3}).
 Within the region, where change of slope occurs
($\mu\simeq-0.6\pm0.005$ and $\mu\simeq-0.425\pm0.005$),
a possibility of three sequential phase transitions of the first
order at change of temperature is observed (phase
diagram in figure~\ref{Fig6}a).
 Near the region of the critical point, a qualitative change
(change of topological type of the phase coexistence
curve and appearance of the triple point are observed)
due to gaussian fluctuations becomes significant when
chemical potential has a value within the mentioned range
(figure~\ref{Fig6}b).

\begin{figure}[hp]
\begin{center}
 \raisebox{-2.3cm}[2.4cm][2.4cm]
 {\epsfxsize 5.cm\epsfbox{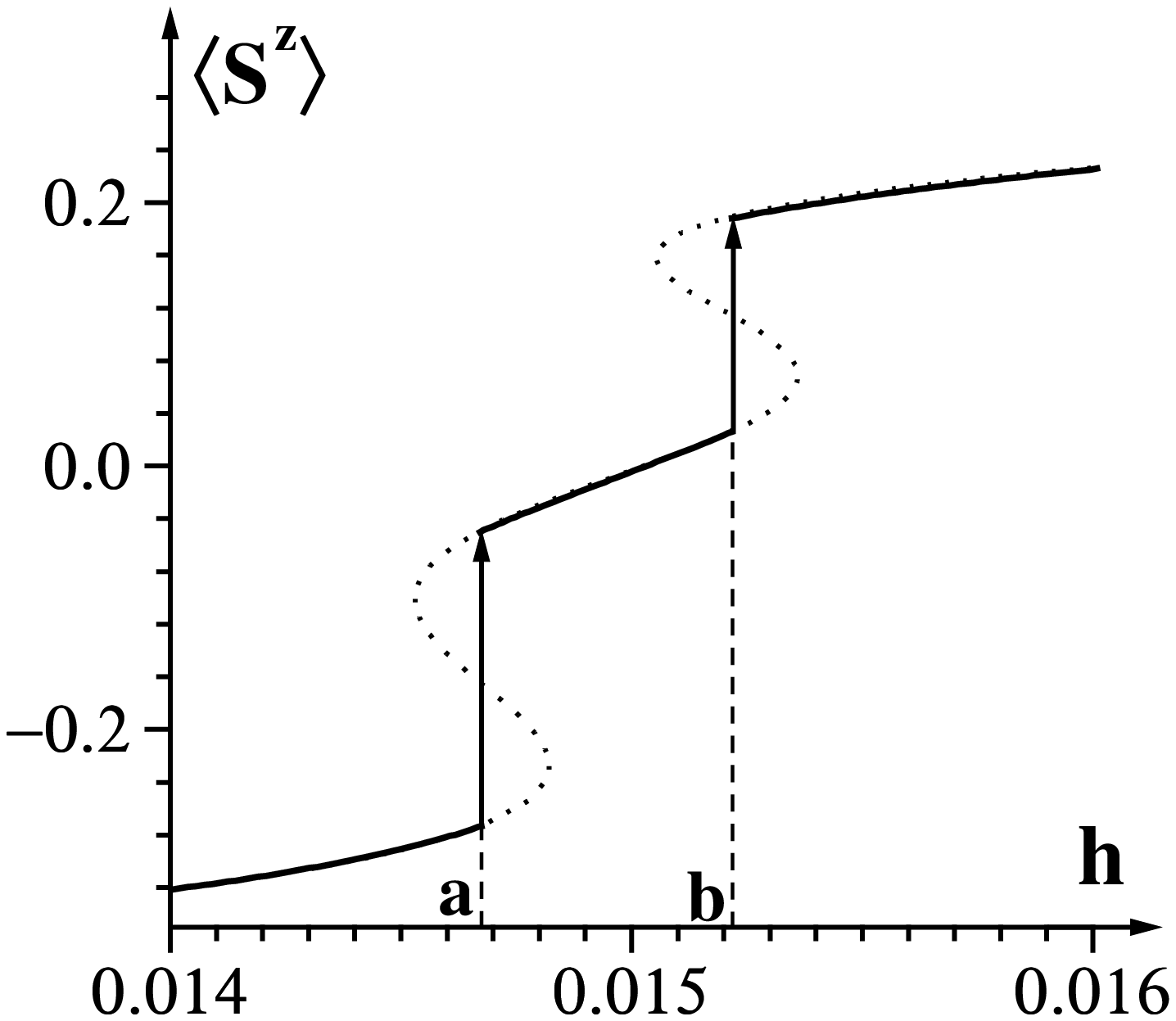}}\; (a)\qquad
 \raisebox{-2.3cm}[2.4cm][2.4cm]
 {\epsfxsize 5.cm\epsfbox{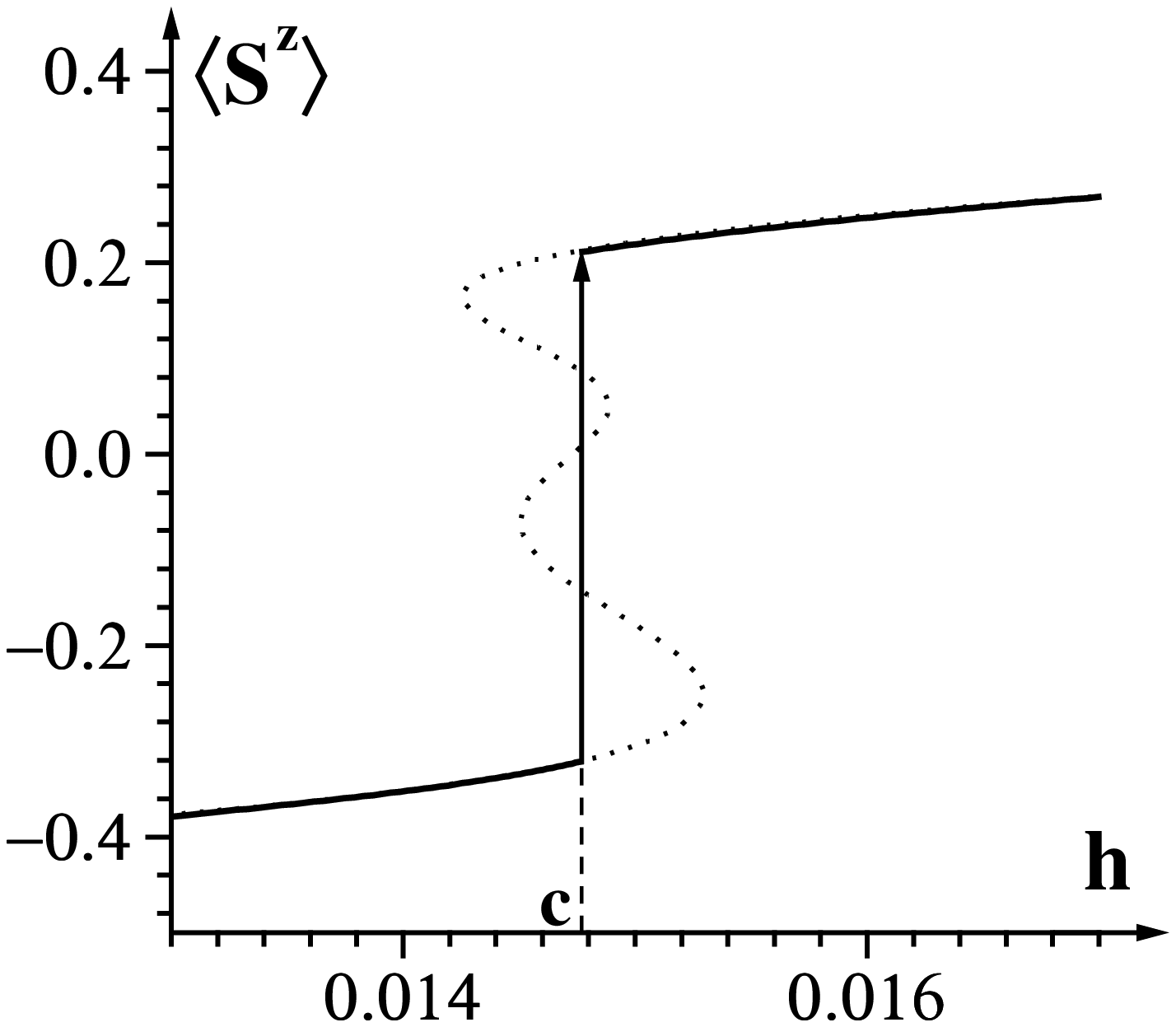}}\; (b)
\end{center}
\caption{Field dependences of pseudospin mean value:
        (a) above the triple point ($T=0.01370$),
        (b) below the triple point ($T=0.01316$)
        ($g=1$, $t_{\bk=0}=0.2$, $\mu=-0.6$).}
 \label{Fig7}
\end{figure}

\begin{figure}[hp]
\begin{center}
 \raisebox{-2.3cm}[2.4cm][2.4cm]
 {\epsfxsize 5.cm\epsfbox{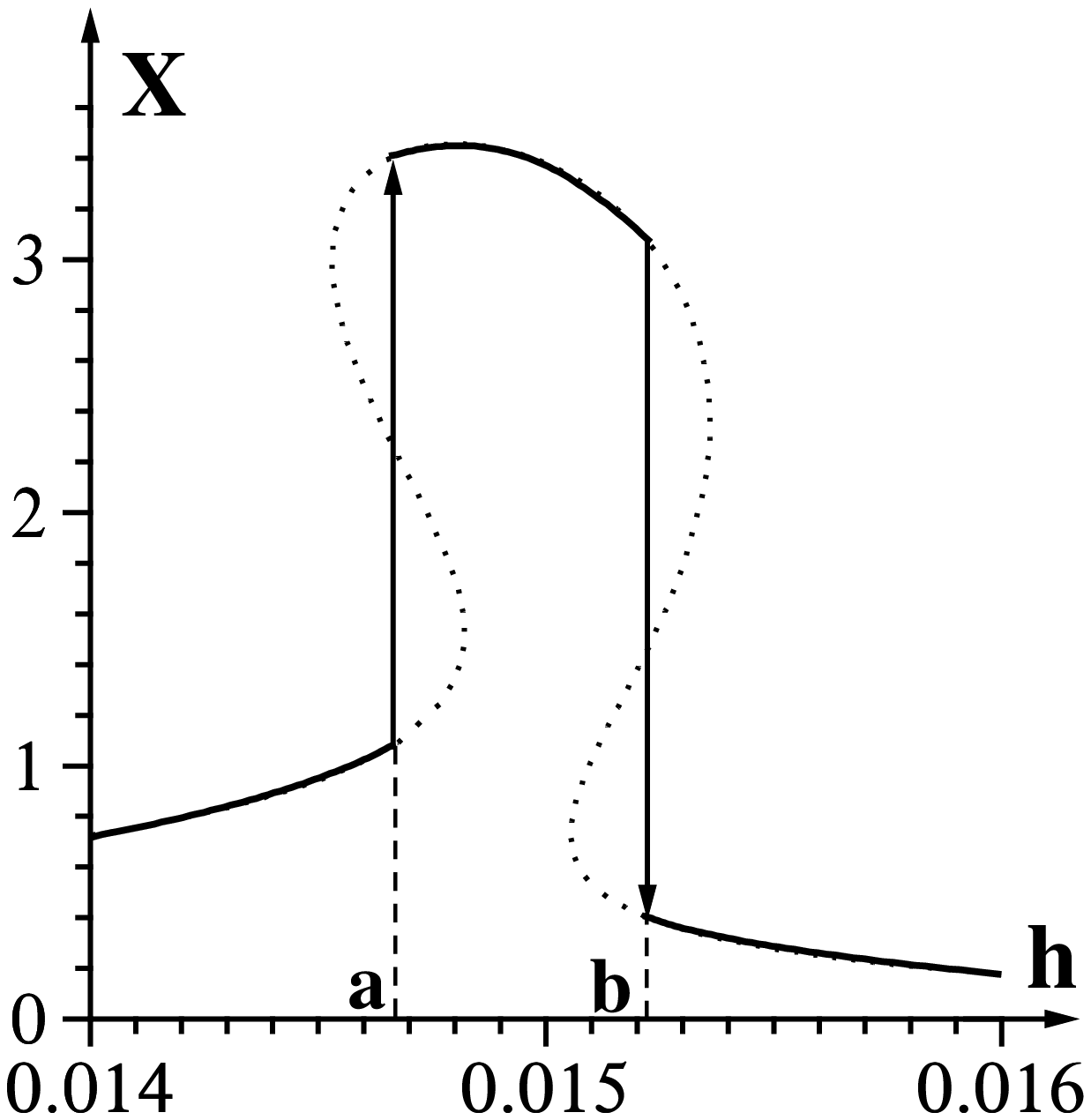}}\; (a)\qquad
 \raisebox{-2.3cm}[2.4cm][2.4cm]
 {\epsfxsize 5.cm\epsfbox{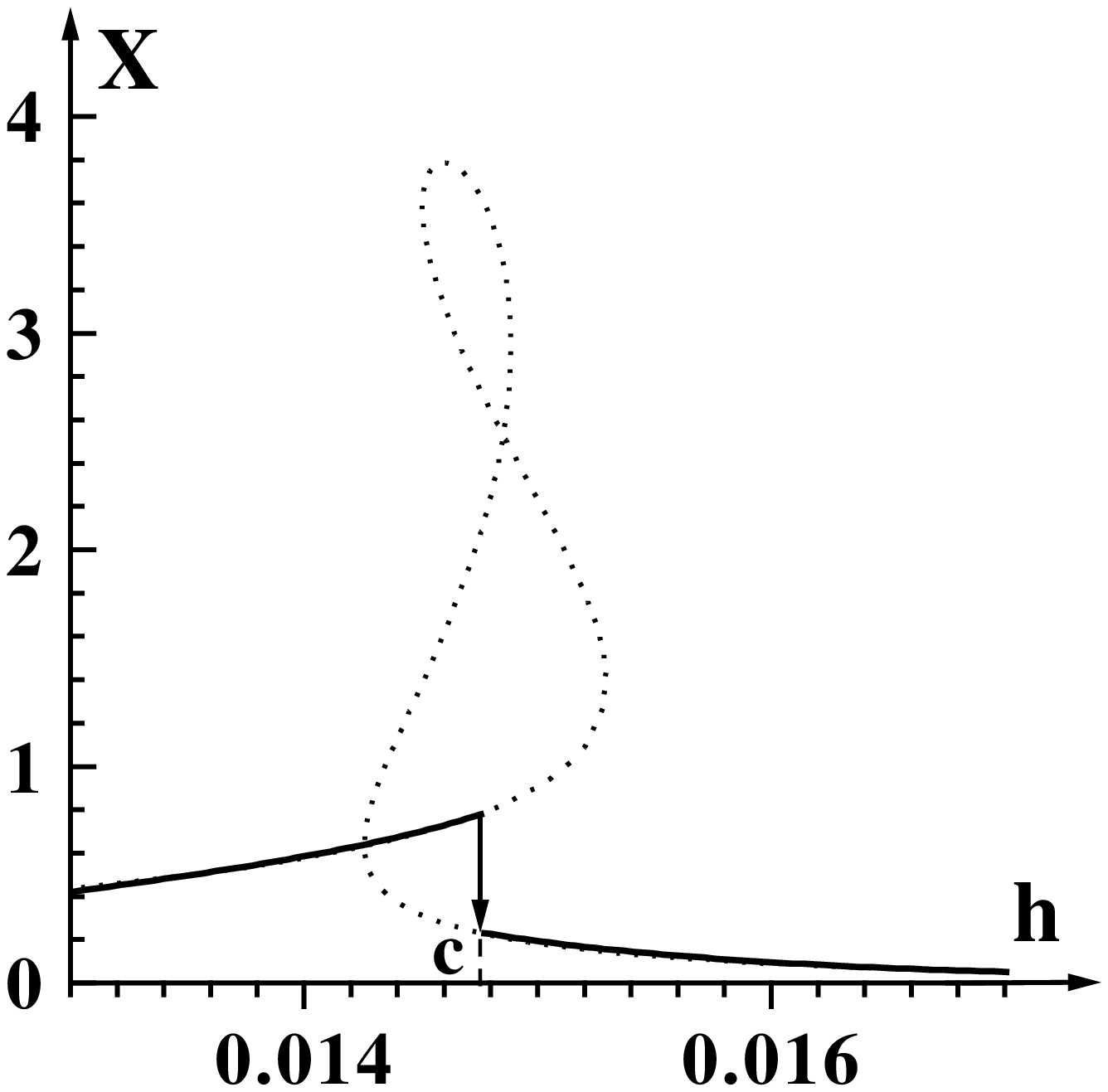}}\; (b)
\end{center}
\caption{Field dependences of r.m.s. fluctuation parameter:
        (a) above the triple point ($T=0.01370$),
        (b) below the triple point ($T=0.01316$)
        ($g=1$, $t_{\bk=0}=0.2$, $\mu=-0.6$).}
 \label{Fig8}
\end{figure}

 The field dependences of pseudospin mean value and r.m.s.
fluctuation parameter in the temperature vicinity of this region
(near the triple point) are presented in figure~\ref{Fig7}
and in figure~\ref{Fig8}
((a) -- above the triple point, (b) -- below the triple point).
\begin{figure}[hp]
\begin{center}
 \raisebox{-2.3cm}[2.4cm][2.4cm]
 {\epsfxsize 5.cm\epsfbox{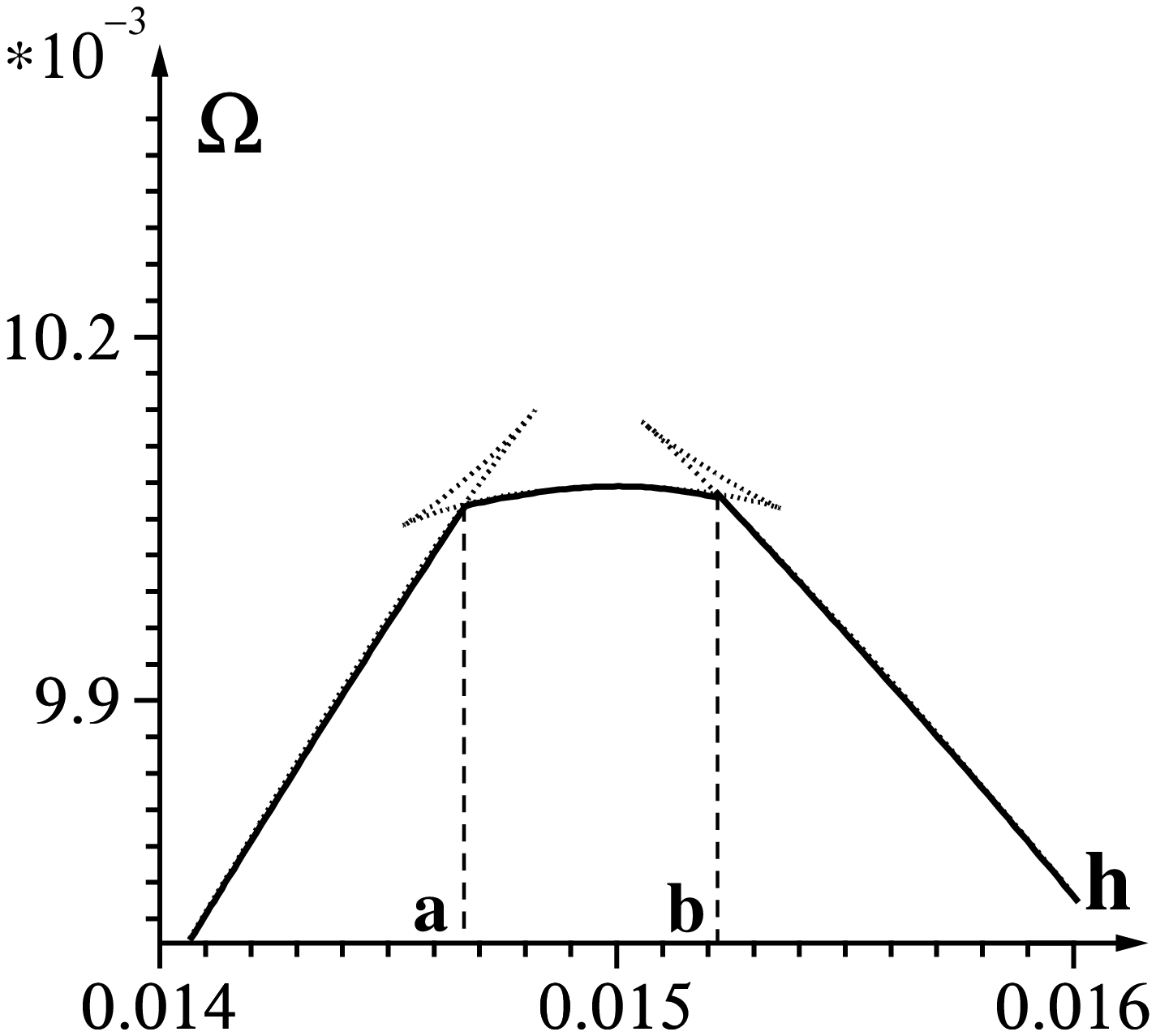}}\; (a)\qquad
 \raisebox{-2.3cm}[2.4cm][2.4cm]
 {\epsfxsize 5.cm\epsfbox{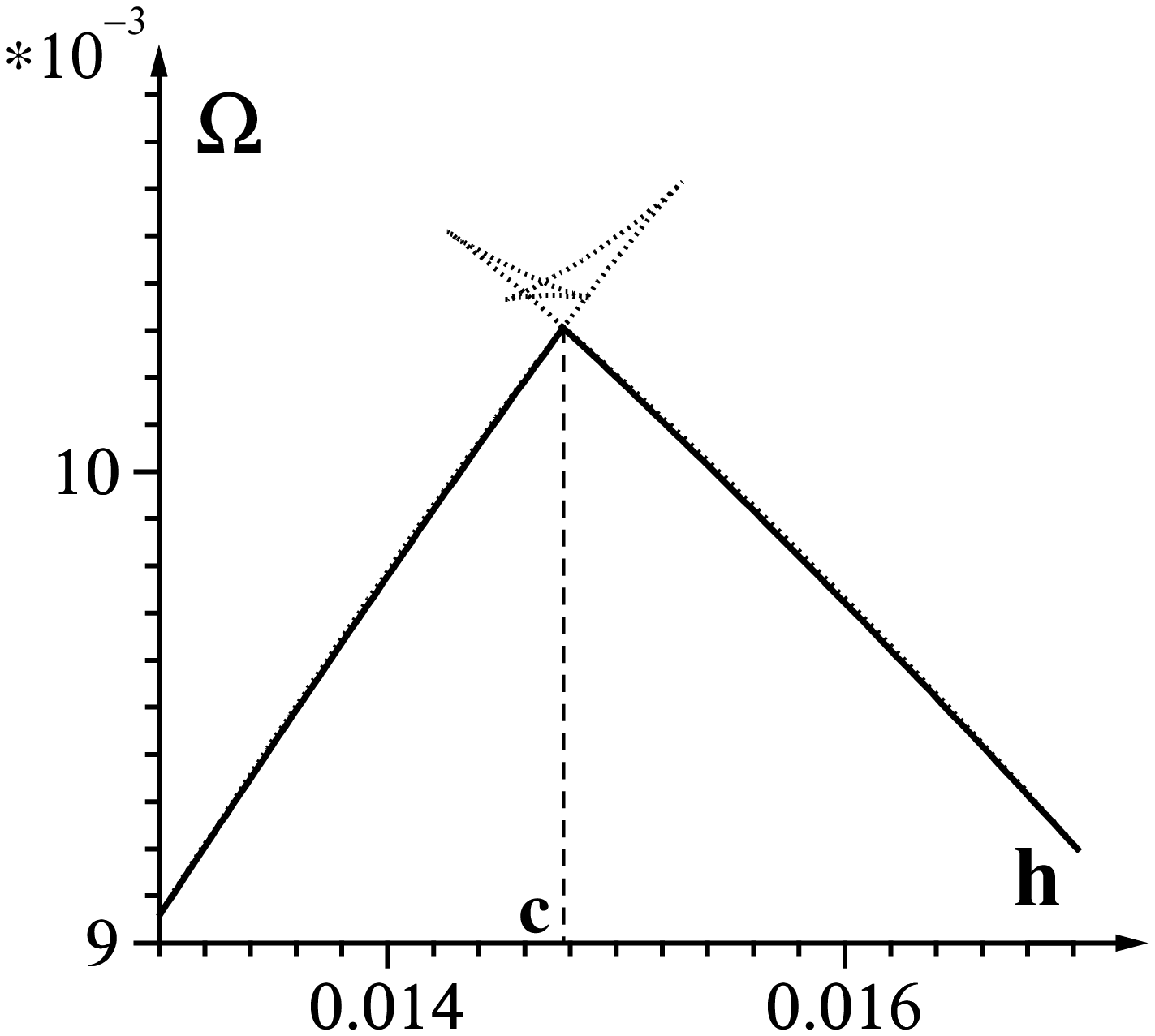}}\; (b)
\end{center}
\caption{Field dependences of grand canonical potential:
        (a) above the triple point ($T=0.01370$),
        (b) below the triple point ($T=0.01316$)
        ($g=1$, $t_{\bk=0}=0.2$, $\mu=-0.6$).}
 \label{Fig9}
\end{figure}
 The phase transition points
(denoted as \textbf{a}, \textbf{b} and \textbf{c} points in figures)
correspond to the crossing points of different branches of
$\Omega (h)$ (figure\ref{Fig9}).

\begin{figure}[htb]
\begin{center}
 \raisebox{-2.3cm}[2.4cm][2.4cm]
 {\epsfxsize 5.cm\epsfbox{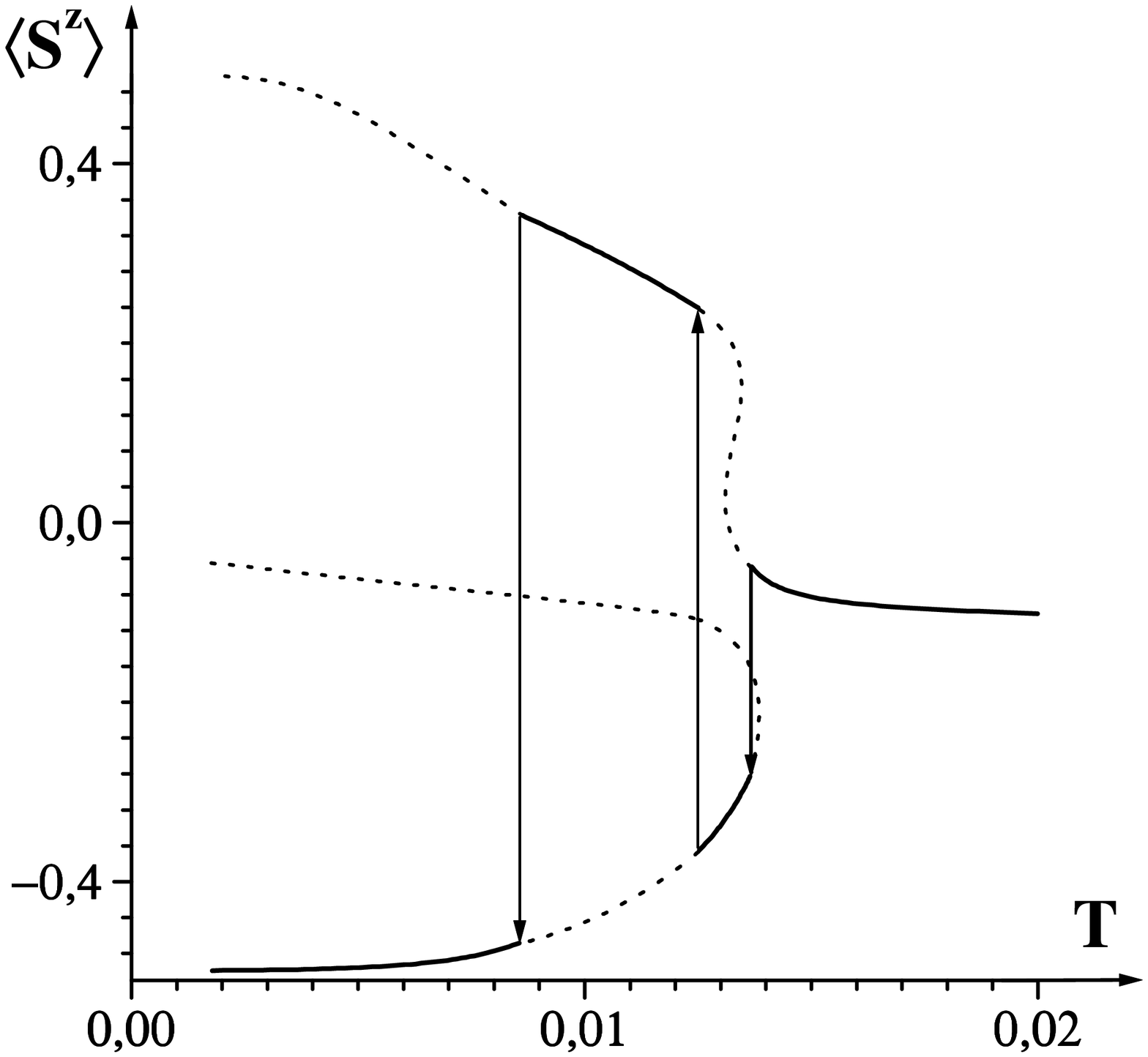}}\; (a)\qquad
 \raisebox{-2.3cm}[2.4cm][2.4cm]
 {\epsfxsize 5.cm\epsfbox{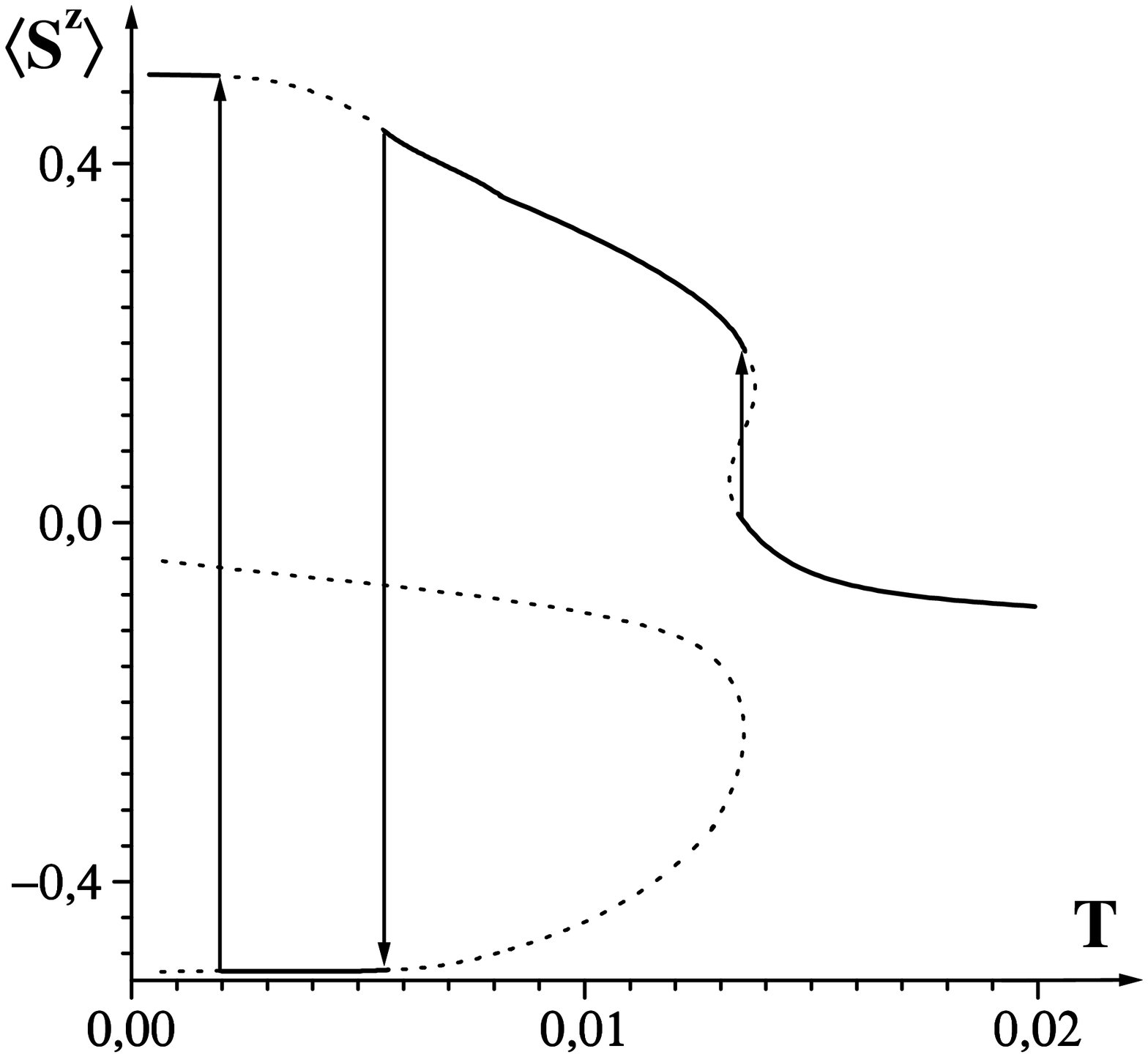}}\; (b)
\end{center}
\caption{Temperature dependences of the pseudospin mean value:
(a) $h=0.01467$,
(b) $h=0.01495$.}
 \label{Fig10}
\end{figure}

 In the figures presented above the case when chemical potential is placed
in the lower subband is presented.
 If the chemical potential is placed in the upper subband our results
transform according to the mentioned above electron-hole symmetry
of the initial Hamiltonian.

\section{Conclusion}

 Self-consistent method, taking into account corrections
due to gaussian fluctuations of effective field in the
self-consistent GRPA scheme, and its simplified version
(analogous to the approximation proposed by Onyszkiewicz
\cite{Onyshkevych1,Onyshkevych2} for spin models) are applied for
investigation of pseudospin-electron model.

 Calculations of the thermodynamic functions have shown that Onyszkiewicz type
approach (taking into account the class of diagrams restricted as
against the usual gaussian fluctuation approach)
does not change qualitatively any of the results obtained
within the gaussian fluctuation approximation.
 A comparison with mean field type approximations (self-consistent GRPA, DMFA)
shows that taking into account of fluctuations is essential in the region of
the critical point and can lead to the qualitative changes
of behavior of thermodynamical functions and the shape of corresponding
phase diagrams for some special value of chemical potential.
 The presented results demonstrate that the quantitative changes due to
fluctuations lower the critical point temperature
and shift the corresponding value of longitudinal field
when chemical potential is placed in electron bands.
 The lowest value of critical temperatures correspond to the Onyszkiewicz type
approximation.

 Preliminary analysis of temperature behaviour of the pseudospin correlator
$\langle S^zS^z\rangle_{\bq}$ (\ref{correlatior})
(with fixed r.m.s. fluctuation parameter)
shows that the high temperature phase become
unstable with respect to the fluctuations with ${\bq}\not =0$
when chemical potential is placed between the electron subbands.
 The maximal temperature of instability is achieved for ${\bq}=(\pi,\pi)$
and indicates a possibility of phase transition into a modulated (chess-board)
phase.
 This result confirms the previously obtained within the framework
of the self-consistent GRPA one \cite{Tabunshchyk2,Tabunshchyk3},
but taking into account of fluctuations lowers value of temperature in
which the instability occurs.

 In this paper we investigated the possible phase transitions in
PEM within the gaussian fluctuation approximation without creation
of super structures ($\bq =0$) and all the presented phase diagrams
concern only this case.
 Presented in our paper method of consideration of gaussian
fluctuations of the effective mean field can be used
to investigation of the influence of fluctuations on
thermodynamic properties of modulated (chess-board) phase
(like it was done in work \cite{Tabunshchyk3} within the framework
of the self-consistent GRPA).
 This issue will be the subject of the further investigation.

\ack

 The author wishes to thank Professor I.V. Stasyuk for his helpful discussions
and for a critical reading of the paper.

%
%

\end{document}